\pdfoutput=1
\newcommand*{\ATLASLATEXPATH}{latex/}
\documentclass[cernpreprint, texlive=2016, UKenglish]{\ATLASLATEXPATH atlasdoc}
\usepackage{\ATLASLATEXPATH atlaspackage}
\usepackage{\ATLASLATEXPATH atlasbiblatex}
\usepackage{\ATLASLATEXPATH atlascontribute}
\usepackage{\ATLASLATEXPATH atlasphysics}
\usepackage{grffile}
\usepackage{tikz}
\usetikzlibrary{positioning}
\def\checkmark{\tikz\fill[scale=0.4](0,.35) -- (.25,0) -- (1,.7) -- (.25,.15) -- cycle;}
\usepackage{\ATLASLATEXPATH atlascover}
\makeatletter
\def\blx@maxline{77}
\makeatother
\graphicspath{{logos/}{figures/}}
\usepackage{rotating}
\usepackage{longtable,lscape}
\usepackage{multirow}                                                                                                 
\usepackage{url}
\usepackage{units}
\usepackage{pifont}

\def\Hp {\ensuremath{H^+}}
\def\Hm {\ensuremath{H^-}}

\def\bbar {\ensuremath{{\overline b}}}
\def\tbar {\ensuremath{{\overline t}}}
\def\bbbar {\ensuremath{{b\bbar}}}
\def\ttbar {\ensuremath{{t\tbar}}}

\def\invfb {\ensuremath{{\mathrm{fb}^{-1}}}}

\def\fb {\ensuremath{{\mathrm{fb}}}}
\def\pb {\ensuremath{{\mathrm{pb}}}}
\def\nb {\ensuremath{{\mathrm{nb}}}}

\newcommand{\LUMI}{36.1\,\invfb }
\newcommand{\LUMIfive}{3.2\,\invfb }
\newcommand{\LUMIsix}{32.9\,\invfb }

\newcommand{\LUMIERROR}{36.1 $\pm$ 0.8~\invfb }
\usepackage{ifthen}

\def\mt{\ensuremath{m_{\mathrm{T}}}}
\def\tauhadvis{\ensuremath{\tau_{\text{had-vis}}}}

\def\sc{\scshape}

\AtlasTitle{Search for charged Higgs bosons decaying via $H^{\pm} \to \tau^{\pm}\nu_{\tau}$ in the $\tau$+jets and $\tau$+lepton final states with 36~\invfb\ of $pp$ collision data recorded at $\sqrt{s} = 13$~\TeV\ with the ATLAS experiment}
\author{The ATLAS Collaboration}

\AtlasRefCode{HIGG-2016-11}

\PreprintIdNumber{CERN-EP-2018-148}

\arXivId{1807.07915}

\AtlasJournal{JHEP}
\AtlasJournalRef{JHEP 09 (2018) 139}
\AtlasDOI{10.1007/JHEP09(2018)139}

\AtlasAbstract{%
Charged Higgs bosons produced either in top-quark decays or in association with a top-quark, 
subsequently decaying via $H^{\pm} \to \tau^{\pm}\nu_{\tau}$, are searched for in \LUMI\ of 
proton--proton collision data at $\sqrt{s}=13$~\TeV\ recorded with the ATLAS detector. Depending 
on whether the top-quark produced together with $H^{\pm}$ decays hadronically or leptonically, 
the search targets $\tau$+jets and $\tau$+lepton final states, in both cases with a hadronically 
decaying $\tau$-lepton. No evidence of a charged Higgs boson is found. For the mass range of 
$m_{H^{\pm}}$ = 90--2000~\GeV, 
upper limits at the 95\% confidence level are set on the production cross-section of the charged 
Higgs boson times the branching fraction $\mathrm{\cal{B}}(H^{\pm} \to \tau^{\pm}\nu_{\tau})$ in 
the range 4.2--0.0025~pb. In the mass range 90--160~\GeV, assuming the Standard Model cross-section 
for \ttbar\ production, this corresponds to upper limits between 0.25\% and 0.031\% for the branching fraction
$\mathrm{\cal{B}}(t\to bH^{\pm}) \times \mathrm{\cal{B}}(H^{\pm} \to \tau^{\pm}\nu_{\tau})$.
}

\hypersetup{pdftitle={ATLAS document},pdfauthor={The ATLAS Collaboration}}

\begin{document}
\maketitle
\tableofcontents

\section{Introduction}
\label{sec:intro}
\begin{figure}[htbp]
  \centering
\subfloat[\label{fig:hplus-diagrams_a}]{\includegraphics[width=0.3\linewidth]{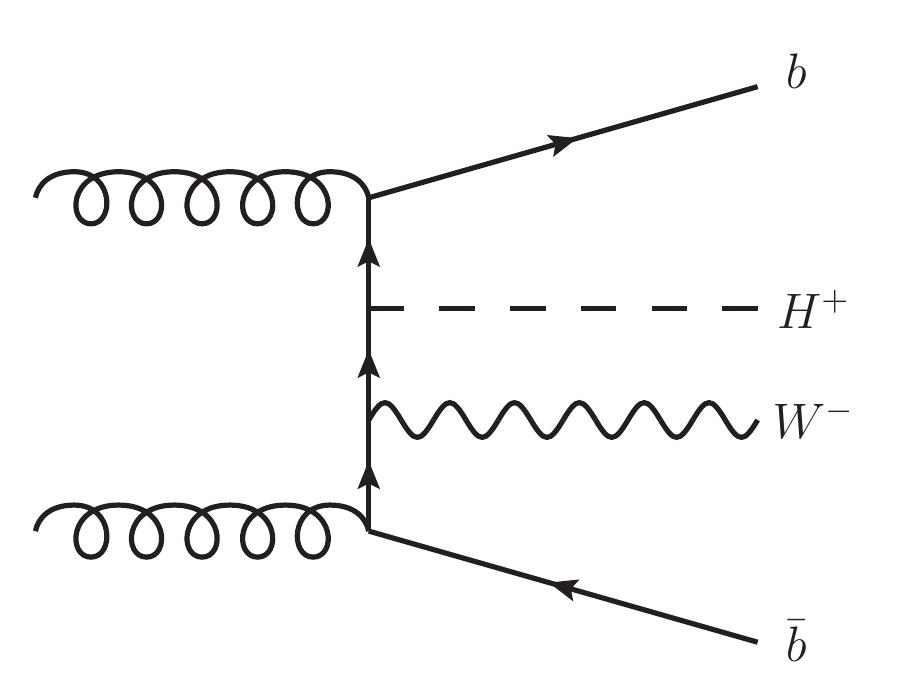}}
\subfloat[\label{fig:hplus-diagrams_b}]{\includegraphics[width=0.3\linewidth]{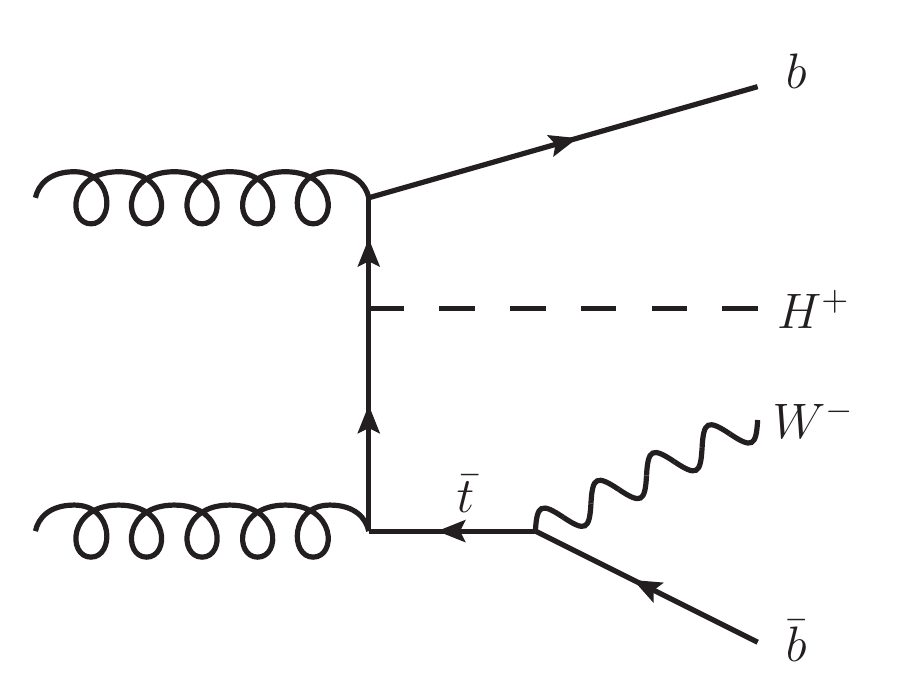}}
\subfloat[\label{fig:hplus-diagrams_c}]{\includegraphics[width=0.3\linewidth]{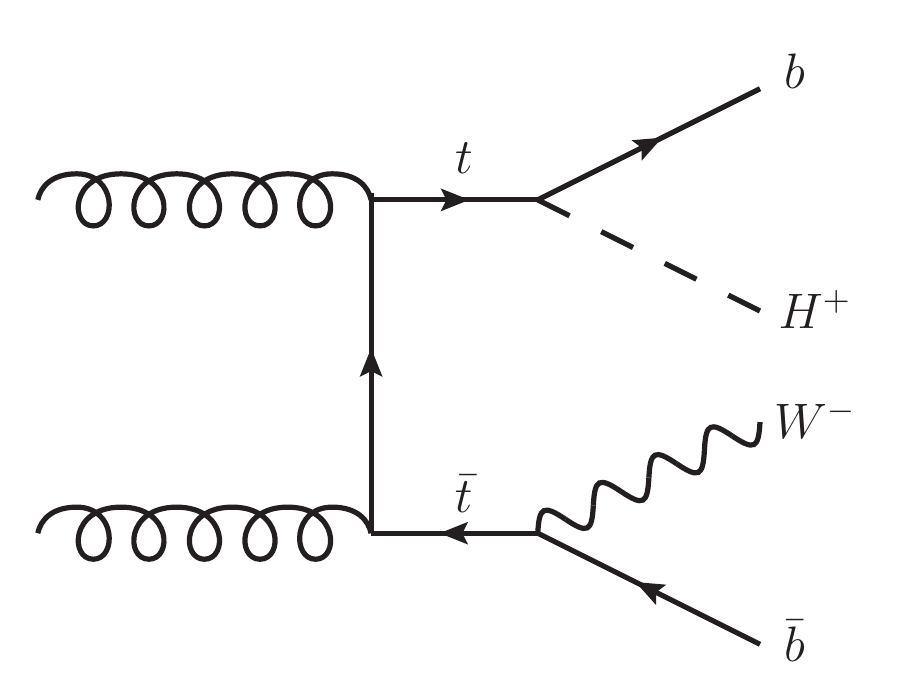}}
\caption{\label{fig:hplus-diagrams} Examples of leading-order Feynman diagrams contributing to the production of 
charged Higgs bosons in $pp$ collisions: (a) non-resonant top-quark production, (b) single-resonant 
top-quark production that dominates at large $\Hp$ masses, (c) double-resonant top-quark production that 
dominates at low $\Hp$ masses. The interference between these three main diagrams becomes 
most relevant in the intermediate-mass region.}
\end{figure}

The discovery of a new particle at the Large Hadron Collider (LHC)~\cite{Evans:2008}
in 2012~\cite{Aad:2012tfa,Chatrchyan:2012xdj}, with a measured mass 
close to 125~\GeV~\cite{Aad:2015zhl}, opens the question of whether this 
is the Higgs boson of the Standard Model (SM), or part of an extended scalar sector. 
Charged Higgs bosons\footnote{In the 
following, charged Higgs bosons are denoted $\Hp$, with the 
charge-conjugate $\Hm$ always implied. Generic symbols are also used for particles 
produced in association with charged Higgs bosons and in their decays.} are predicted in several 
extensions of the SM that add a second doublet~\cite{Djouadi:2005gj,Branco:2011iw} 
or triplets~\cite{Senjanovic:1975rk,CiezaMontalvo:2006zt,Gunion:1989ci,Georgi:1985nv}
to its scalar sector. In CP-conserving Two-Higgs-Doublet Models (2HDMs), the properties
of the charged Higgs boson depend on its mass, the mixing angle $\alpha$ of the neutral 
CP-even Higgs bosons and the ratio of the vacuum expectation values of the two Higgs 
doublets ($\tan\beta$). Although the search for charged Higgs bosons presented in this paper 
is performed in a model-independent manner, results are interpreted in the framework of the 
hMSSM benchmark scenario~\cite{Djouadi:2013uqa,Djouadi:2015jea}, which is a type-II 2HDM, 
where down-type quarks and charged leptons couple to one Higgs doublet, while up-type quarks 
couple to the other.
For \Hp\ masses below the top-quark mass ($m_{\Hp} < m_{\text{top}}$), the main production
mechanism is through the decay of a top-quark, $t \to b\Hp$, in a double-resonant top-quark 
production. In this mass range, the decay $\Hp \to \tau\nu$ usually dominates in a type-II 2HDM, 
although $\Hp \to cs~\mbox{and}~cb$ may also become sizeable at low $\tan\beta$. For \Hp\ masses 
above the top-quark mass ($m_{\Hp} > m_{\text{top}}$), the leading production mode is
$gg \to tb\Hp$ (single-resonant top-quark production). Close to the alignment limit, 
i.e.\ at $\cos(\beta-\alpha) \simeq 0$, the dominant decay is $\Hp \to tb$; however the 
branching fraction of $\Hp \to \tau\nu$ can reach 10--15\% at large values of $\tan\beta$ 
in a type-II 2HDM. In the intermediate-mass region ($m_{\Hp} \simeq m_{\text{top}}$), 
accurate theoretical predictions recently became available for the non-resonant top-quark 
production~\cite{Degrande:2016hyf}, which now allows a dedicated comparison of the $\Hp$ models 
with data near the top-quark mass. Figure~\ref{fig:hplus-diagrams} illustrates the main production 
modes for charged Higgs bosons in proton--proton ($pp$) collisions.

The ATLAS and CMS Collaborations searched for charged Higgs bosons in $pp$ collisions at 
$\sqrt{s}$ = 7--8~\TeV, probing the mass range below the top-quark mass with the
$\tau\nu$~\cite{Aad:2012tj,Aad:2012rjx,Aad:2014kga,Chatrchyan:2012vca,Khachatryan:2015qxa}
and $cs$~\cite{Aad:2013hla,Khachatryan:2015uua} decay modes, as
well as the mass range above the top-quark mass with the $\tau\nu$
and $tb$ decay modes~\cite{Aad:2014kga,Khachatryan:2015qxa,Aad:2015typ}. 
More recently, using 3.2\,\invfb\ of data collected at $\sqrt{s} = 13$~\TeV, searches 
for charged Higgs bosons heavier than $m_{\text{top}}$ were performed by ATLAS in the
$\tau\nu$ decay mode~\cite{Aaboud:2016dig}.
In addition, $\Hp \to WZ$ was searched for in the vector-boson-fusion production mode 
at 8~\TeV\ by the ATLAS Collaboration~\cite{Aad:2015nfa} and at 13~\TeV\ by the CMS 
Collaboration~\cite{Sirunyan:2017sbn}. No evidence of charged Higgs bosons was found 
in any of these searches. 
The ATLAS and CMS Collaborations also searched for neutral scalar resonances 
decaying to a $\tau\tau$ pair~\cite{Aaboud:2017sjh,Sirunyan:2018zut}, 
to which the hMSSM is also sensitive in some regions of its parameter space. 

This paper describes a search for charged Higgs bosons using the
$\Hp \to \tau\nu$ decay, with a subsequent hadronic decay of the
$\tau$-lepton (referred to as $\tau_{\text{had}}$), in the mass range
90--2000~\GeV, including the intermediate-mass region. 
Depending on the assumption made for the decay mode of the $W$ boson originating
from the top-quark produced together with \Hp, two channels are targeted: 
$\tau_{\text{had}}$+jets if the $W$ boson decays into a $q\bar{q}'$ pair, 
or $\tau_{\text{had}}$+lepton if the $W$ boson decays into an electron 
or muon and at least one neutrino (directly or via a leptonically decaying $\tau$-lepton). 
The data used for this analysis are from $pp$ collisions at $\sqrt{s}=13$~\TeV, 
collected with the ATLAS experiment at the LHC in 2015 and 2016,
corresponding to integrated luminosities of \LUMIfive\ and \LUMIsix,
respectively. In Section~\ref{sec:data_samples}, the data and simulated
samples are summarised. In Section~\ref{sec:object_reconstruction}, the
reconstruction of physics objects is described. The analysis
strategy and event selection are discussed in Section~\ref{sec:event_selection}.
Section~\ref{sec:background_estimation} describes the data-driven estimation
of backgrounds with misidentified $\tau$ objects.
A discussion of the systematic uncertainties and a description of the statistical analysis 
used to derive exclusion limits on the production of a charged Higgs boson decaying via 
$H^+ \rightarrow \tau\nu$ are presented in Sections~\ref{sec:systematic_uncertainties} 
and~\ref{sec:results}, respectively. Finally, a summary is given in Section~\ref{sec:conclusion}.

\section{Data and samples of simulated events}
\label{sec:data_samples}
The ATLAS experiment~\cite{Aad:2008zzm} is a multipurpose
detector with a forward-backward symmetric cylindrical geometry with respect to the LHC
beam-axis.\footnote{The ATLAS experiment uses a right-handed coordinate system with its
origin at the nominal interaction point (IP) in the centre of the detector and the $z$-axis
along the beam pipe. The $x$-axis points from the IP to the centre of the LHC ring, and the
$y$-axis points upward. Cylindrical coordinates $(r,\phi)$ are used in the transverse plane,
$\phi$ being the azimuthal angle around the $z$-axis. The pseudorapidity is defined in terms
of the polar angle $\theta$ as $\eta = -\ln[\tan(\theta/2)]$. Transverse momenta are computed
from the three-momenta $\vec{p}$ as $\pt = |\vec{p}|\sin\theta$.
The distance in the $\eta$-$\phi$ space is commonly referred to as
$\Delta R = \sqrt{(\Delta \eta)^2 + (\Delta \phi)^2}$.}
The innermost layers of ATLAS consist of tracking detectors in the pseudorapidity range
$|\eta|<2.5$, including the insertable B-layer~\cite{Capeans:2010jnh,ATLAS:2012rar} installed for 
Run-2. This inner detector is surrounded by a thin superconducting solenoid that provides
a 2~T axial magnetic field. It is enclosed by the electromagnetic and hadronic calorimeters which
cover $|\eta|<4.9$. The outermost layers consist of an external muon spectrometer within
$|\eta|<2.7$, incorporating three large toroidal magnet assemblies. A two-level 
trigger~\cite{Aaboud:2016leb,ATL-DAQ-PUB-2017-001} reduces the event rate to a 
maximum of 1~kHz for offline data storage.

The dataset used in this analysis, collected during stable beam conditions and with all ATLAS 
subsystems fully operational, corresponds to an integrated luminosity of \LUMIERROR, derived 
following a methodology similar to that detailed in Ref.~\cite{Aaboud:2016hhf}. 
Data-quality criteria are applied in order to remove events where reconstructed jets are 
consistent with noise in the calorimeter or non-collision backgrounds~\cite{Aaboud:2017jcu}.

Signal events with $H^+ \to \tau\nu$ were generated with
{\sc MadGraph5\_aMC@NLO}~\cite{Alwall:2014hca} in three distinct mass regions:
\begin{itemize}
\item In the mass range below the top-quark mass (90--150~\GeV), \ttbar\ events with one 
top-quark decaying into a charged Higgs boson and a $b$-quark were generated at leading 
order (LO). Contributions from events in which both top-quarks decay via $t \to b\Hp$ and 
from single-top-quark events with a subsequent decay $t \to b\Hp$ are negligible, hence 
not simulated.
\item In the intermediate-mass region (160--180~\GeV), the full process $pp \to H^+Wbb$ was 
generated at LO. 
\item In the mass range above the top-quark mass (200--2000~\GeV), $H^+$ production 
in association with a top-quark was simulated at next-to-leading order (NLO).
\end{itemize}
The interference between the Feynman diagrams of Figure~\ref{fig:hplus-diagrams} is taken into account in the intermediate-mass region, where it is most relevant. 
In all cases, the matrix-element generator was interfaced to {\sc Pythia v8.186}~\cite{Sjostrand:2007gs}
for the simulation of the parton shower and the underlying event, with the A14~\cite{ATL-PHYS-PUB-2014-021} 
set of tuned parameters (tune). The NNPDF2.3~\cite{Ball:2012cx} parton distribution 
function (PDF) sets were used for the matrix-element generation and the parton shower.

The SM background processes include the production of \ttbar\ pairs, single top-quarks,
$W$+jets, $Z/\gamma^*$+jets and electroweak gauge-boson pairs ($WW/WZ/ZZ$), as well as
multi-jet events. The \ttbar\ events constitute the main background in the low- and
intermediate-mass \Hp\ searches, while multi-jet events dominate for large charged Higgs
boson masses. All backgrounds arising from a quark- or gluon-initiated jet misidentified as 
a hadronically decaying $\tau$-lepton are estimated with a data-driven method, described 
in Section~\ref{sec:background_estimation}.

The generation of \ttbar\ events used the 
{\sc Powheg-Box v2}~\cite{Nason:2004rx,Frixione:2007vw,Alioli:2010xd} generator, 
with the CT10~\cite{Lai:2010vv} PDF set in the matrix-element calculations. 
Single-top-quark events were generated in the $Wt$-, $t$- and $s$-channels using the 
{\sc Powheg-Box v1} generator, with the CT10 PDF set in the $Wt$- and $s$-channels or 
the fixed 4-flavour CT10f4~\cite{Dulat:2015mca} PDF set in the $t$-channel. 
For all processes above, top-quark spin correlations 
are preserved ({\sc MadSpin}~\cite{Artoisenet:2012st} was used for top-quark decays in 
the $t$-channel). The parton shower and the underlying event were
simulated using {\sc Pythia v6.428}~\cite{Sjostrand:2006za} with the 
CTEQ6L1~\cite{Pumplin:2002vw} PDF set and the corresponding Perugia 
2012~\cite{Skands:2010ak} tune. The top-quark mass was set to 172.5~\GeV.
The sample of \ttbar\ events is normalised to the next-to-next-to-leading-order
(NNLO) cross-section, including soft-gluon resummation to next-to-next-to-leading-logarithm
(NNLL) order~\cite{Czakon:2011xx}. The normalisation
of the sample of single-top-quark events uses an approximate calculation at NLO in QCD
for the $s$- and $t$-channels~\cite{Kant:2014oha,Aliev:2010zk} and an NLO+NNLL calculation 
for the $Wt$-channel~\cite{Kidonakis:2010ux,Kidonakis:2013zqa}.

Events containing a $W$ or $Z$ boson with associated jets were simulated using
{\sc Sherpa v2.2.1}~\cite{Gleisberg:2008ta} together with the NNPDF3.0NNLO~\cite{Ball:2014uwa} 
PDF set. Matrix elements were calculated for up to two partons at NLO and four partons at LO using 
{\sc Comix}~\cite{Gleisberg:2008fv} and {\sc OpenLoops}~\cite{Cascioli:2011va}, and they were merged 
with the {\sc Sherpa} parton shower~\cite{Schumann:2007mg} according to the 
ME+PS@NLO prescription~\cite{Hoeche:2012yf}. The $W/Z$+jets events are 
normalised to the NNLO cross-sections calculated using
{\sc FEWZ}~\cite{Anastasiou:2003ds,Melnikov:2006kv,Gavin:2010az}.
Diboson processes ($WW$, $WZ$ and $ZZ$) were simulated at NLO using the {\sc Powheg-Box v2}
generator, interfaced to the {\sc Pythia v8.186} parton shower model. The CT10nlo PDF
set was used for the hard-scatter process, while the CTEQL1 PDF set was used for the
parton shower. The non-perturbative effects were modelled using the AZNLO~\cite{Aad:2014xaa}
tune.
{\sc EvtGen v1.2.0}~\cite{Lange:2001uf} was used for the properties of bottom- and charm-hadron decays, 
except in samples generated with {\sc Sherpa}. All simulated events were overlaid with additional
minimum-bias events generated with {\sc Pythia v8.186} using the A2~\cite{ATLAS:2012uec}
tune and the MSTW2008LO~\cite{Martin:2009iq} PDF set to simulate the effect of multiple $pp$ 
collisions per bunch crossing (pile-up). Simulated events were then weighted to have the same 
distribution of the number of collisions per bunch crossing as the data. All signal
and background events were processed through a simulation~\cite{Aad:2010ah} of the detector geometry
and response based on {\sc Geant4}~\cite{Agostinelli:2002hh} and they are reconstructed using the same
algorithms as the data.

\section{Physics object reconstruction}
\label{sec:object_reconstruction}
The search reported in this paper makes use of most of the physics objects 
reconstructed in ATLAS: charged leptons (electrons and muons), jets (including 
those compatible with the hadronisation of $b$-quarks or hadronic decays of 
$\tau$-leptons) and missing transverse momentum. These physics objects are 
detailed below.

Electrons are reconstructed by matching clustered energy deposits in 
the electromagnetic calorimeter to a track reconstructed in the inner 
detector~\cite{ATLAS:2016iqc}. They are required to have $p_{\text{T}}> 20~\GeV$ 
and $|\eta|<2.47$ (the transition region between the
barrel and end-cap calorimeters, $1.37<|\eta|<1.52$, is excluded).
In the $\tau_{\text{had}}$+jets ($\tau_{\text{had}}$+lepton) channel, 
electrons must satisfy a loose (tight) identification criterion based 
on a likelihood discriminant.

Muon candidates are required to contain matching inner-detector and
muon-spectrometer tracks~\cite{Aad:2016jkr}, as well as to have $\pt > 20~\GeV$
and $|\eta| < 2.5$. The final muon tracks are re-fitted using the complete
track information from both detector systems. In the $\tau_{\text{had}}$+jets 
($\tau_{\text{had}}$+lepton) channel, muons must satisfy a loose (tight) identification 
criterion.

In order to ensure that electrons (muons) originate from the primary vertex, defined as 
the vertex with the highest sum of the $\pt^2$ of its associated tracks, the track associated with the lepton is 
required to have a longitudinal impact parameter and a transverse impact parameter significance 
that fulfil, respectively, $|z_{0}\sin\theta| < 0.5$~mm and $|d_{0}/\sigma(d_{0})| < 5~(3)$. 
In order to reduce contamination by leptons from hadron decays or photon conversion, isolation requirements are applied. 
The calorimeter-based isolation relies on energy deposits within a cone of size $\Delta R = 0.2$ around 
the electron or muon, while the track-based isolation uses a variable cone size starting at $\Delta R = 0.2$ 
for electrons or $\Delta R = 0.3$ for muons, and then decreasing as $\pt$ increases. The efficiency of the 
calorimeter- and track-based isolation requirements are, respectively, 96\% and 99\% in the 
$\tau_{\text{had}}$+lepton channel (looser isolation requirements are applied in the 
$\tau_{\text{had}}$+jets channel, with efficiencies of 99\%).

Jets are reconstructed from energy deposits in the calorimeters using the anti-$k_t$ 
algorithm~\cite{Cacciari:2008gp} implemented in the {\sc FastJet} package~\cite{Cacciari:2011ma} 
with a radius parameter value of \mbox{$R = 0.4$}. Jets are corrected for pile-up energy and 
calibrated using energy- and $\eta$-dependent corrections~\cite{Aaboud:2017jcu}. 
Only jets with a transverse momentum $\pt > 25~\GeV$ and within $|\eta |<2.5$ are considered in 
the following. A multivariate technique (jet vertex tagger) that allows identification
and selection of jets originating from the hard-scatter interaction through the
use of tracking and vertexing information is applied to jets with $\pt < 60~\GeV$
and $|\eta|<2.4$~\cite{jvt}. In order to identify jets containing $b$-hadrons (referred 
to as $b$-jets in the following), an algorithm is used, which combines impact parameter 
information with the explicit identification of secondary and tertiary vertices within 
the jet into a multivariate discriminant~\cite{Aad:2015ydr,ATL-PHYS-PUB-2016-012}. Operating 
points are defined by a single threshold in the range of discriminant output values and are chosen 
to provide a specific $b$-jet efficiency in simulated \ttbar\ events. The 70\% working point is 
used in this analysis. It has rejection factors of 13, 56 and 380 against $c$-jets, hadronic 
$\tau$ decays and jets from light quarks or gluons, respectively. 

The detection and reconstruction of hadronically decaying $\tau$-leptons is seeded by anti-$k_t$ jets
depositing a transverse energy $\ET > 10~\GeV$ in the calorimeter~\cite{Aad:2014rga} and
with one or three associated tracks reconstructed in the inner detector within a cone 
of $\Delta R = 0.2$ around the axis of the object associated with the visible decay products 
of the hadronically decaying $\tau$-leptons. Referred to as \tauhadvis\ in the following, these 
objects are required to have a visible transverse momentum ($\pt^{\tau}$) of at least 30~\GeV\ and 
to be within $|\eta| < 2.3$ (the transition region between the barrel and end-cap calorimeters,
$1.37<|\eta|<1.52$, is excluded). In order to distinguish $\tauhadvis$ candidates from quark- 
or gluon-initiated jets, a boosted decision tree (BDT) is used, separately for candidates with 
one and three charged-particle tracks~\cite{ATL-PHYS-PUB-2015-045,ATLAS-CONF-2017-029}.
For the nominal definition of $\tauhadvis$ candidates, a working point with an identification 
efficiency of 75\% (60\%) for 1-prong (3-prong)  hadronic $\tau$ decays in $Z \rightarrow \tau\tau$ 
events is chosen, corresponding to rejection factors of 30--80 (200--1000) against quark- and 
gluon-initiated jets in multi-jet events. An additional likelihood-based veto is used to reduce 
the number of electrons misidentified as \tauhadvis\ candidates, providing a constant 95\% efficiency 
as a function of \tauhadvis\ $\pt$ and $\eta$, as well as a background rejection of 20--200, 
depending on $\eta$.

When several objects overlap geometrically, the following procedure is applied.
First, any $\tauhadvis$ object is removed if found within $\Delta R = 0.2$ of 
either an electron or a muon with loose identification criteria and with a 
transverse momentum above 20~\GeV\ or 7~\GeV, respectively. Then, any 
electron sharing an inner-detector track with a muon is discarded. 
Next, jets within $\Delta R = 0.2$ of an electron or muon are removed, unless they are 
$b$-tagged, have a large $\pt$ relative to that of the lepton and, in the case of an 
overlap with a muon, a high track multiplicity. Then, electrons and muons are removed 
if found within $\Delta R = 0.4$ of a remaining jet. Finally, jets are discarded if they 
are within $\Delta R = 0.2$ of the highest-\pt\ $\tauhadvis$ candidate.

The magnitude of the missing transverse momentum, \met~\cite{Aaboud:2018tkc},
is reconstructed from the negative vector sum of transverse momenta of
reconstructed and fully calibrated objects, with an additional term that 
is calculated from inner-detector tracks that are matched to the primary vertex 
and not associated with any of the selected objects.

\section{Analysis strategy}
\label{sec:event_selection}
This paper describes a search for a charged Higgs boson decaying via 
$H^+ \to \tau\nu$ in topologies where it is produced either in top-quark decays or in association 
with a top-quark. Depending on whether the top-quark produced with the $H^+$ decays hadronically 
or semileptonically, two channels are targeted: $\tauhadvis$+jets or $\tauhadvis$+lepton, 
respectively. The corresponding signal regions are described below.

\subsection*{Event selection in the $\tauhadvis$+jets channel}
\label{sec:event_taujets}

The analysis of the $\tauhadvis$+jets channel is based on events accepted by an $\met$
trigger with a threshold at 70, 90 or 110 \GeV, depending on the data-taking period 
and thereby accounting for different pile-up conditions. 
The efficiency of these triggers is measured in data and used to reweight the simulated events,
with the same method as in Ref.~\cite{Aaboud:2016dig}. At least one vertex with two or more 
associated tracks with $\pt > 400~\MeV$ is required, and the following event selection
criteria are applied:
\begin{itemize}
\item at least one $\tauhadvis$ candidate with $\pt^\tau > 40~\GeV$;
\item no loose leptons (electron or muon) with $\pt > 20~\GeV$;
\item at least three jets with $\pt > 25~\GeV$, of which at least one is $b$-tagged;
\item $\met > 150~\GeV$;
\item $\mt > 50$~GeV.
\end{itemize}

Here, the transverse mass $\mt$ of the highest-$\pt$ $\tauhadvis$ candidate and
$\met$ is defined as,
\begin{equation}
  \label{eq:mt}
  \mt=\sqrt{ 2 \pt^\tau  \ET^{\mathrm{miss}} (1-\cos \Delta\phi_{\tau,\text{miss}}) },
\end{equation}
where $\Delta\phi_{\tau,\text{miss}}$ is the azimuthal angle between the $\tauhadvis$
candidate and the direction of the missing transverse momentum.

\subsection*{Event selection in the $\tauhadvis$+lepton channel}
\label{sec:event_taulep}

The $\tauhadvis$+electron and $\tauhadvis$+muon sub-channels are based on events accepted 
by single-lepton triggers. Triggers for electrons or muons with low $E_{\text{T}}$ or $\pt$ 
thresholds respectively (24--26 \GeV\ depending on the data-taking period, for both 
the electrons and muons) and isolation requirements are combined in a logical OR with
triggers having higher $(E_{\text{T}},\pt)$ thresholds (60--140~\GeV\ for electrons, 50~\GeV\ for
muons) and looser isolation or identification requirements in order to maximise the efficiency.
Following the same vertex requirement as in the $\tauhadvis$+jets channel, events are 
selected as follows:
\begin{itemize}
\item exactly one lepton matched to the single-lepton trigger object, with 
$\pt > 30~\GeV$. Depending on whether the lepton is an electron or a muon, 
two sub-channels, $\tauhadvis$+electron and $\tauhadvis$+muon, are considered;
\item exactly one $\tauhadvis$ candidate with $\pt^\tau > 30~\GeV$ and an electric charge opposite 
to that of the lepton;
\item at least one $b$-tagged jet with $\pt > 25~\GeV$;
\item $\met > 50~\GeV$.
\end{itemize}

\subsection*{Multivariate discriminant}

Following the event selections above, kinematic variables that differentiate between 
the signal and backgrounds are identified and combined into a multivariate discriminant. 
The output score of BDTs is then used in order to separate the \Hp\ signal from the 
SM background processes. The training of the BDTs is performed using the FastBDT~\cite{Keck:2017gsv} 
library via the TMVA toolkit~\cite{Hocker:2007ht}. 

The simulated signal samples are divided into five \Hp\ mass bins chosen to ensure 
that within each bin both the kinematic distributions of the input variables and the 
event topology are similar. 
The mass bins used in both channels are 90--120~\GeV,
130--160~\GeV\ (in that case, an additional signal sample with a 160~\GeV\ \Hp\ arising 
solely from top-quark decays is used), 160--180~\GeV, 200--400~\GeV\ and
500--2000~\GeV. All available \Hp\ signal samples corresponding to a given mass 
bin are normalised to the same event yield and combined into one inclusive signal 
sample. The BDTs are trained separately for $\tauhadvis$+jets and $\tauhadvis$+lepton 
events, and depending on whether the leading \tauhadvis\ candidate has one or three 
associated tracks. The variables entering the BDT training differ for the two types 
of final states considered in this search, and they are summarised in
Table~\ref{Tab:BDT}. If there is more than one \tauhadvis\ candidate or more than one 
$b$-tagged jet, the object that has the largest $\pt$ is considered in the BDT input variables.

At low \Hp\ masses, the kinematics of the $t \to bH^+$ and $t \to bW$ decay products 
are similar. In that case, the polarisation of the $\tau$-lepton is employed as
a discriminating variable: in the main SM background processes, the \tauhadvis\ object
originates from a vector-boson decay, whereas it is generated in the decay of a
scalar particle in the case of \Hp\ signal~\cite{Roy:1999xw}. 
The polarisation of the \tauhadvis\ candidates 
can be measured by the asymmetry of energies carried by the charged and neutral
pions from the 1-prong $\tau$-lepton decay, measured in the laboratory frame.
For this purpose, the variable $\Upsilon$ is introduced~\cite{Aad:2012cia}:
\begin{equation*}
\Upsilon = \frac{\et^{\pi^{\pm}}-\et^{\pi^{0}}}{\et^{\tau}} \approx 2 {\pt^{\tau{\text{-track}}} \over \pt^{\tau}}-1.
\end{equation*}
It is defined for all \tauhadvis\ candidates with only one associated track,
and $\pt^{\tau{\text{-track}}}$ is the transverse momentum of that track. 
For \Hp\ masses in the range 90--400~\GeV, the BDT training is performed
separately for events with a selected 1- or 3-prong \tauhadvis\ object,
and $\Upsilon$ is included in the BDT discriminant for events where
\tauhadvis\ has only one associated track. While $\Upsilon$ is one of the 
most discriminating input variables of the BDT at low \Hp\ masses, the 
importance of other kinematic variables in the BDT training becomes much 
greater at large \Hp\ masses, in particular the three variables entering 
the computation of the tranvserse mass, i.e.\ $\met$, $\pt^{\tau}$ and 
$\Delta\phi_{\tau,\text{miss}}$. Hence, for the mass range 500--2000~\GeV,
the BDT discriminant does not contain the variable $\Upsilon$ and is thus 
inclusive in the number of tracks associated with the \tauhadvis\ candidate.

\begin{table}[ht]
\begin{center}
\caption{List of kinematic variables used as input to the BDT in the
$\tauhadvis$+jets and $\tauhadvis$+lepton channels. Here, $\ell$ refers 
to the selected lepton (electron or muon). $\Delta \phi_{X,\,\text{miss}}$ 
denotes the difference in azimuthal angle between a reconstructed object $X$ 
($X = \tauhadvis,\,b\text{-jet},\,\ell$) and the direction of the missing transverse
momentum. The variable $\Upsilon$ is related to the polarisation of 
the $\tau$-lepton and is only defined for 1-prong \tauhadvis\
candidates. Hence, for \Hp\ masses in the range 90--400~\GeV, where the
variable $\Upsilon$ is used, the BDT training is performed separately for
events with a selected 1- or 3-prong \tauhadvis\ candidate. In the mass
range 500--2000~\GeV, $\Upsilon$ is not used, hence the BDT training is
inclusive in number of tracks associated with \tauhadvis\ candidates.\\}
\begin{tabular}{l|c|c}
BDT input variable & $\tauhadvis$+jets & $\tauhadvis$+lepton \\
\hline \hline
$\met$ & \checkmark & \checkmark \\
$\pt^{\tau}$ & \checkmark & \checkmark \\
$\pt^{b\text{-jet}}$ & \checkmark & \checkmark \\
$\pt^{\ell}$ & & \checkmark \\
$\Delta \phi_{\tauhadvis,\,\text{miss}}$ & \checkmark & \checkmark \\
$\Delta \phi_{b\text{-jet},\,\text{miss}}$ & \checkmark & \checkmark \\
$\Delta \phi_{\ell,\,\text{miss}}$ & & \checkmark \\
$\Delta R_{\tauhadvis,\,\ell}$ & & \checkmark \\
$\Delta R_{b\text{-jet},\,\ell}$ & & \checkmark \\
$\Delta R_{b\text{-jet},\,\tauhadvis}$ & \checkmark & \\
$\Upsilon$ & \checkmark & \checkmark \\
\end{tabular}
\label{Tab:BDT}
\end{center}
\end{table}

\section{Background modelling}
\label{sec:background_estimation}
The dominant background processes are categorised according to the object that gives rise to the 
identified \tauhadvis\ candidate. Simulation is used to estimate backgrounds in which \tauhadvis\ 
arises from a hadronically decaying $\tau$-lepton, electron or muon at the event-generator level; 
however, in the case of \ttbar\ events, the normalisation is obtained from a fit to the data. If 
\tauhadvis\ arises from a quark- or gluon-initiated jet, a data-driven method is employed to 
estimate the corresponding background.

\subsection*{Data-driven fake-factor method}

Background processes where a quark- or gluon-initiated jet is reconstructed and selected as a \tauhadvis\
candidate are estimated from data. For this purpose, an anti-\tauhadvis\ selection is defined by requiring 
that the \tauhadvis\ candidate does not satisfy the BDT-based identification criteria of the nominal selection. 
Meanwhile, a loose requirement on the \tauhadvis\ BDT output score is maintained in order to ensure that the 
relative fractions of gluon- and quark-initiated jets mimicking \tauhadvis\ candidates are similar in  
the signal region and the corresponding anti-\tauhadvis\ region. Then, a fake factor (FF) is defined 
as the ratio of the number of jets reconstructed as \tauhadvis\ candidates that pass the nominal 
\tauhadvis\ selection to the number that pass the anti-\tauhadvis\ selection in a given control region (CR):
\begin{equation*}
\mbox{FF} = \frac{N^{\text{CR}}_{\tauhadvis}}{N^{\text{CR}}_{\text{anti-\tauhadvis}}}.
\end{equation*}
The actual computation of FFs is described later in the text. Events are selected with the nominal criteria 
described in Section~\ref{sec:event_selection}, except that an inverted identification criterion for the 
\tauhadvis\ candidate is required.  In this sample, the $\tau_{\text{had}}$ contribution is subtracted using 
simulated events in which a $\tau_{\text{had}}$ at generator level fulfils the anti-\tauhadvis\ criterion. 
The resulting number of events is $N_{\text{anti-\tauhadvis}}$. Then, the number of events with a misidentified 
\tauhadvis\ candidate ($N_{\text{fakes}}^{\tauhadvis}$) is derived from the subset of anti-\tauhadvis\ candidates 
as follows:
\begin{equation*}
N_{\text{fakes}}^{\tauhadvis} = \displaystyle{\sum_i N_{\text{anti-\tauhadvis}}(i) \times \text{FF}(i)},
\end{equation*}
where the index $i$ refers to each bin in the parameterisation of the FF, in terms of $\pt^\tau$
and number of associated tracks.

\begin{figure}[htbp]
\centering
\subfloat[\label{fig:mm:ff:maina}]{\includegraphics[width=0.5\linewidth]{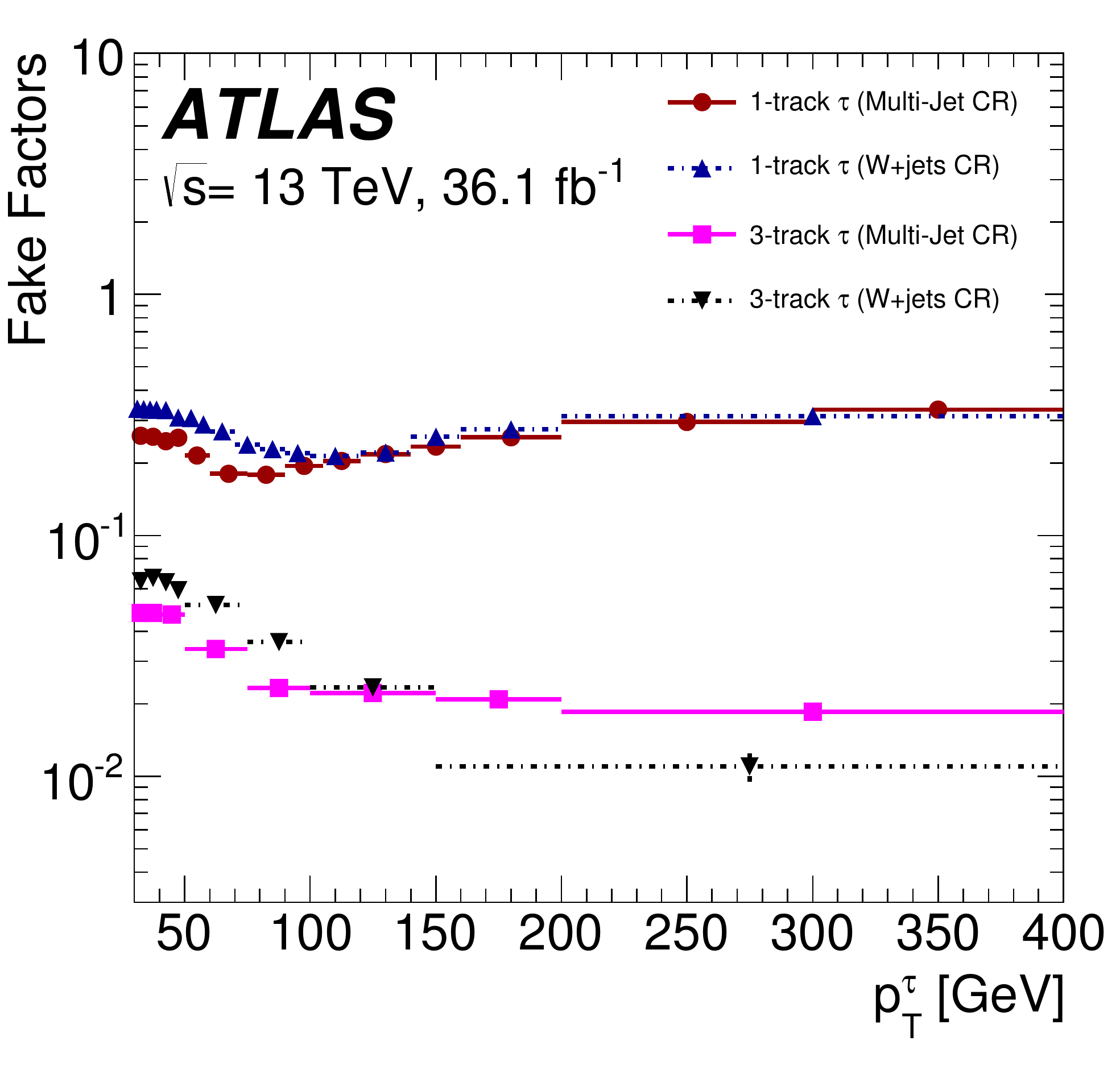}}
\subfloat[\label{fig:mm:ff:mainb}]{\includegraphics[width=0.5\linewidth]{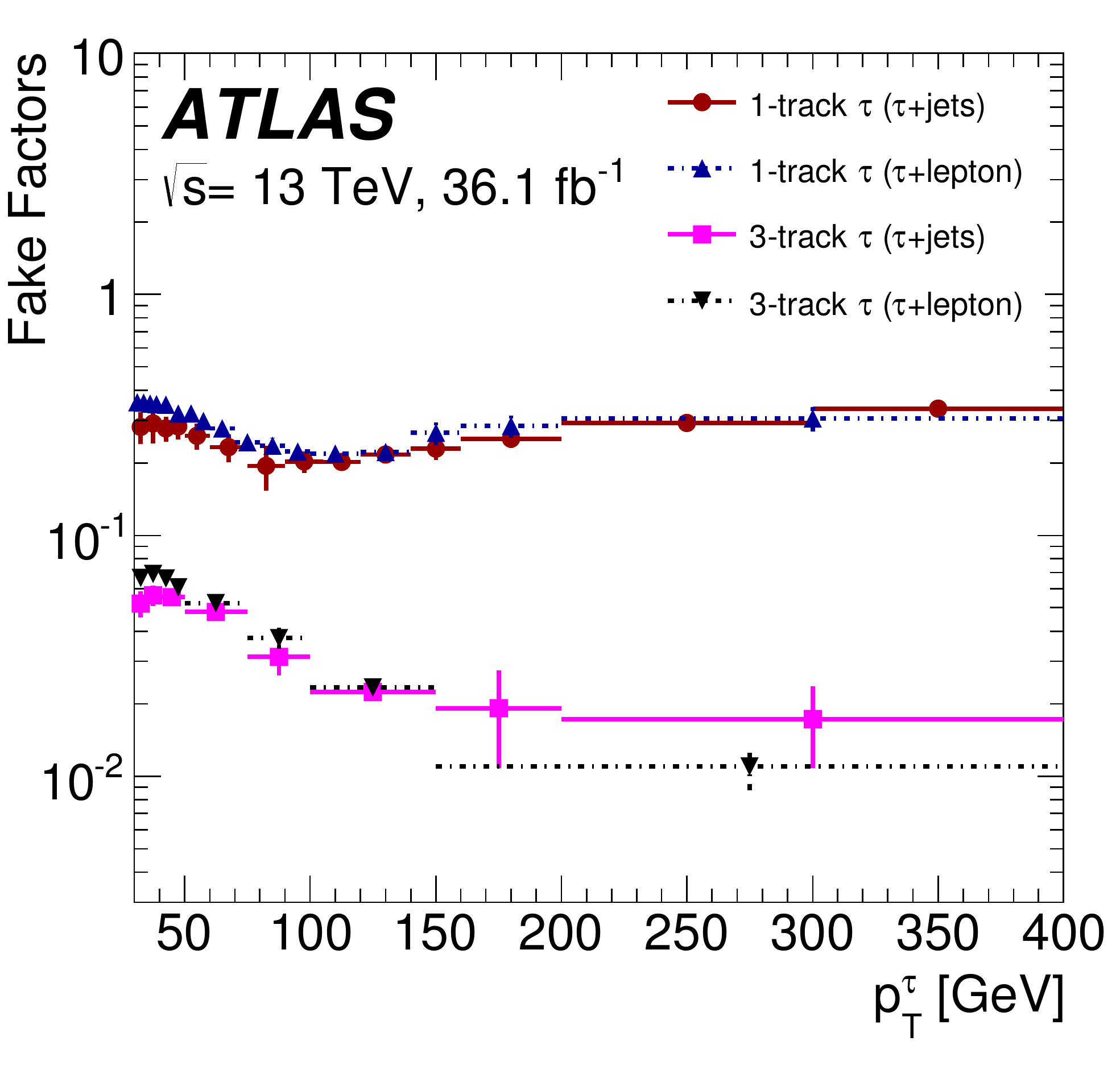}}
\caption{\label{fig:mm:ff:main}Fake factors parameterised as a function of $\pt^\tau$ and the number
of charged $\tau$ decay products (two categories: 1-prong and 3-prong), as obtained
(a) in the multi-jet and $W$+jets CRs, (b) after reweighting by $\alpha_{\text{MJ}}$
in the \tauhadvis+jets and \tauhadvis+lepton channels. Details about the computation 
of fake factors in the different control regions are given in the text.
The errors shown represent the statistical uncertainty in a given $\pt^\tau$ bin (a)
and with additional systematic uncertainties obtained from the combination in a given
$\pt^\tau$ bin (b).}
\end{figure}

In order to account for different sources of misidentified \tauhadvis\ candidates in the
signal region and the corresponding anti-\tauhadvis\ region, FFs are measured in two control regions
of the data with different fractions of quark- and gluon-initiated jets, and then combined. A first
control region with a significant fraction of gluon-initiated jets (referred to as the multi-jet CR) is 
defined by applying the same event selection as for the \tauhadvis+jets channel, but with a $b$-jet 
veto and $\MET < 80~\GeV$. Such events are collected using a combination of multi-jet triggers instead 
of the $\MET$ trigger. A second control region enriched in quark-initiated jets (referred to as the $W$+jets 
CR) is defined by applying the same event selection as for the combined \tauhadvis+lepton channel, but with a 
$b$-jet veto, no requirement on $\MET$ and the requirement $60~\GeV < \mt(\ell,\met) < 160~\GeV$, where the
transverse mass of the lepton and the missing transverse momentum is computed by replacing
the \tauhadvis\ candidate by a lepton in Eq.~(\ref{eq:mt}). The FFs measured in these two
control regions are shown in Figure~\ref{fig:mm:ff:maina}.

In the anti-\tauhadvis\ regions corresponding to the nominal event selections of 
Section~\ref{sec:event_selection}, the fractions of \tauhadvis\ candidates arising 
from quark- and gluon-initiated jets are then measured using a template-fit approach, 
based on variables that are sensitive to the differences in quark- and gluon-fractions between 
these two types of jets. For 3-prong \tauhadvis\ candidates, the \tauhadvis\ BDT output score is used
as a template. For 1-prong \tauhadvis\ candidates, the so-called \tauhadvis\ width is used, defined as:
\begin{equation*}
w_\tau=\frac{\sum \left[ \pt^\mathrm{track}\times\Delta R\left(\tauhadvis,\mathrm{track}\right) \right]}{\sum \pt^\mathrm{track}},
\end{equation*}
for tracks satisfying $\Delta R\left(\tauhadvis,\mathrm{track}\right)<0.4$.
Two binned templates, denoted $f_\text{multi-jet}$ and $f_{W\text{+jets}}$, are obtained in the
multi-jet and $W$+jets control regions defined above, respectively. Each corresponds to a
linear combination of templates of gluon- and quark-initiated jets, where the fraction of
gluon-initiated jets is by construction larger in the multi-jet control region. Then, a
linear combination of the two templates is defined as
$f(x|\alpha_{\text{MJ}})=\alpha_{\text{MJ}}\times f_\text{multi-jet}(x) +   
\left(1-\alpha_{\text{MJ}}\right)\times f_{W\text{+jets}}(x)$ with a free parameter
$\alpha_{\text{MJ}}$ (here, $f(x)$ is the \tauhadvis\ width or BDT score). This
linear combination is fitted to the normalised distribution of the \tauhadvis\ width
or BDT score measured in the anti-\tauhadvis\ regions corresponding to the nominal 
event selections, by varying $\alpha_{\text{MJ}}$ and separately
minimising a $\chi^2$-function in every bin of the fake factors, separately in the 
$\tauhadvis$+jets and $\tauhadvis$+lepton channels. From the best-fit values of 
$\alpha_{\text{MJ}}$, combined fake factors are then given by:
\begin{equation*}
\text{FF}^\mathrm{comb}(i) = \alpha_{\text{MJ}}(i) \times \text{FF}^\text{multi-jet}(i) +
\left[ 1-\alpha_{\text{MJ}}(i) \right] \times \text{FF}^{W\text{+jets}}(i),
\end{equation*}
where the index $i$ refers to each bin in the parameterisation of the FF, and where 
$\text{FF}^\text{multi-jet}$ and $\text{FF}^{W\text{+jets}}$ indicate the FF calculated 
in the two respective control regions. 
The combined FFs, used in the $\tauhadvis$+jets and $\tauhadvis$+lepton channels, 
are shown in Figure~\ref{fig:mm:ff:mainb}.

The data-driven method as described does not correctly predict the shape of 
$\Upsilon$ measured in the signal region. Indeed, the distribution of $\Upsilon$ is found to be different 
for \tauhadvis\ and anti-\tauhadvis\ candidates, because this variable is strongly correlated
with the leading-track momentum fraction, which is one of the input variables 
used for the identification of \tauhadvis\ candidates. On the other hand, $\Upsilon$ shows no 
correlation with any of the other variables used as input to the final BDT discriminant. 
Hence, an inverse transform sampling method~\cite{Smirnov} can be employed in order to 
model the shape of $\Upsilon$ for misidentified \tauhadvis\ candidates in the 
signal regions. In the control regions where FFs are measured, 
cumulative distribution functions $F(\Upsilon)$ are calculated from the shapes of $\Upsilon$, 
obtained separately for \tauhadvis\ and anti-\tauhadvis\ candidates. Then, in the signal 
regions, the shape of $\Upsilon$ predicted for \tauhadvis\ candidates is derived from that 
measured for anti-\tauhadvis\ candidates, as follows:
\begin{equation*}
\Upsilon_{\tauhadvis} = F_{\tauhadvis}^{-1}(F_{\text{anti-}\tauhadvis}(\Upsilon)),
\end{equation*}
where $F^{-1}$ stands for the inverse of the cumulative distribution function. This procedure 
is only applied to 1-prong objects, since $\Upsilon$ is not used in the training of the final 
BDT discriminant for 3-prong \tauhadvis\ candidates.

\subsection*{Validation of the background modelling}

\begin{figure}[htbp]
\centering
\subfloat[\label{fig:taulepBDTscoreDLT:a}]{\includegraphics[width=0.45\linewidth]{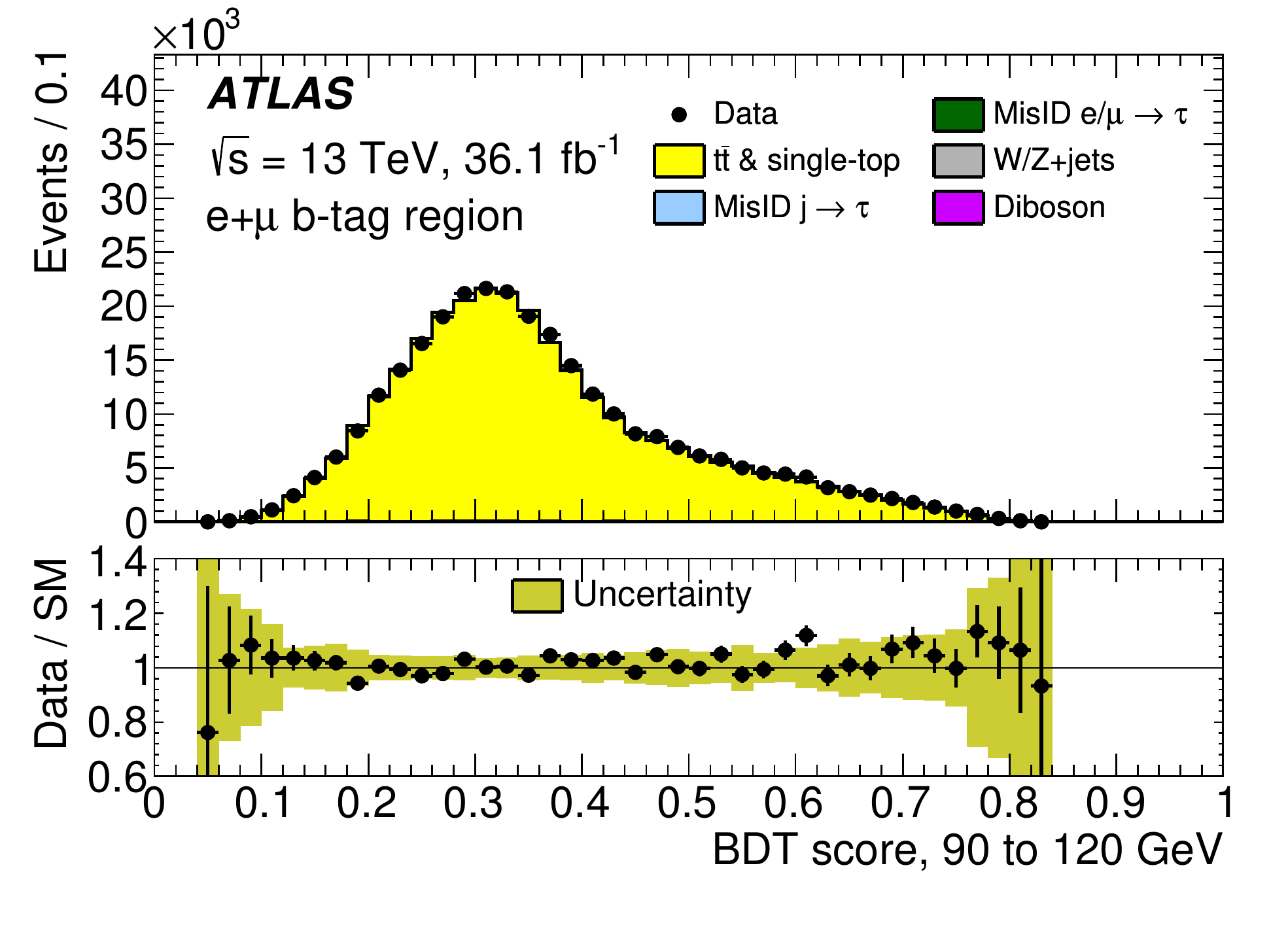}}
\subfloat[\label{fig:taulepBDTscoreDLT:b}]{\includegraphics[width=0.45\linewidth]{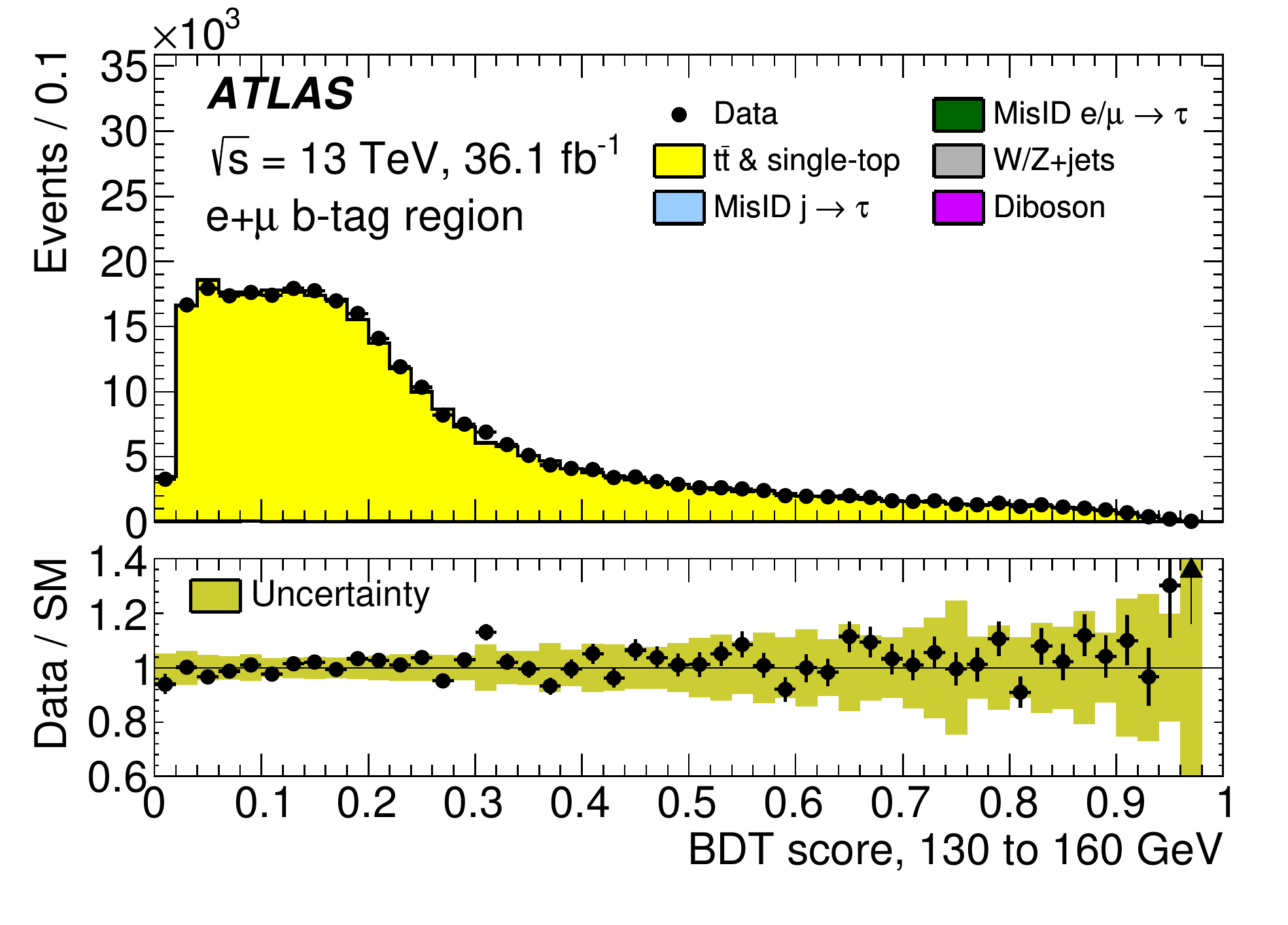}} \\
\subfloat[\label{fig:taulepBDTscoreDLT:c}]{\includegraphics[width=0.45\linewidth]{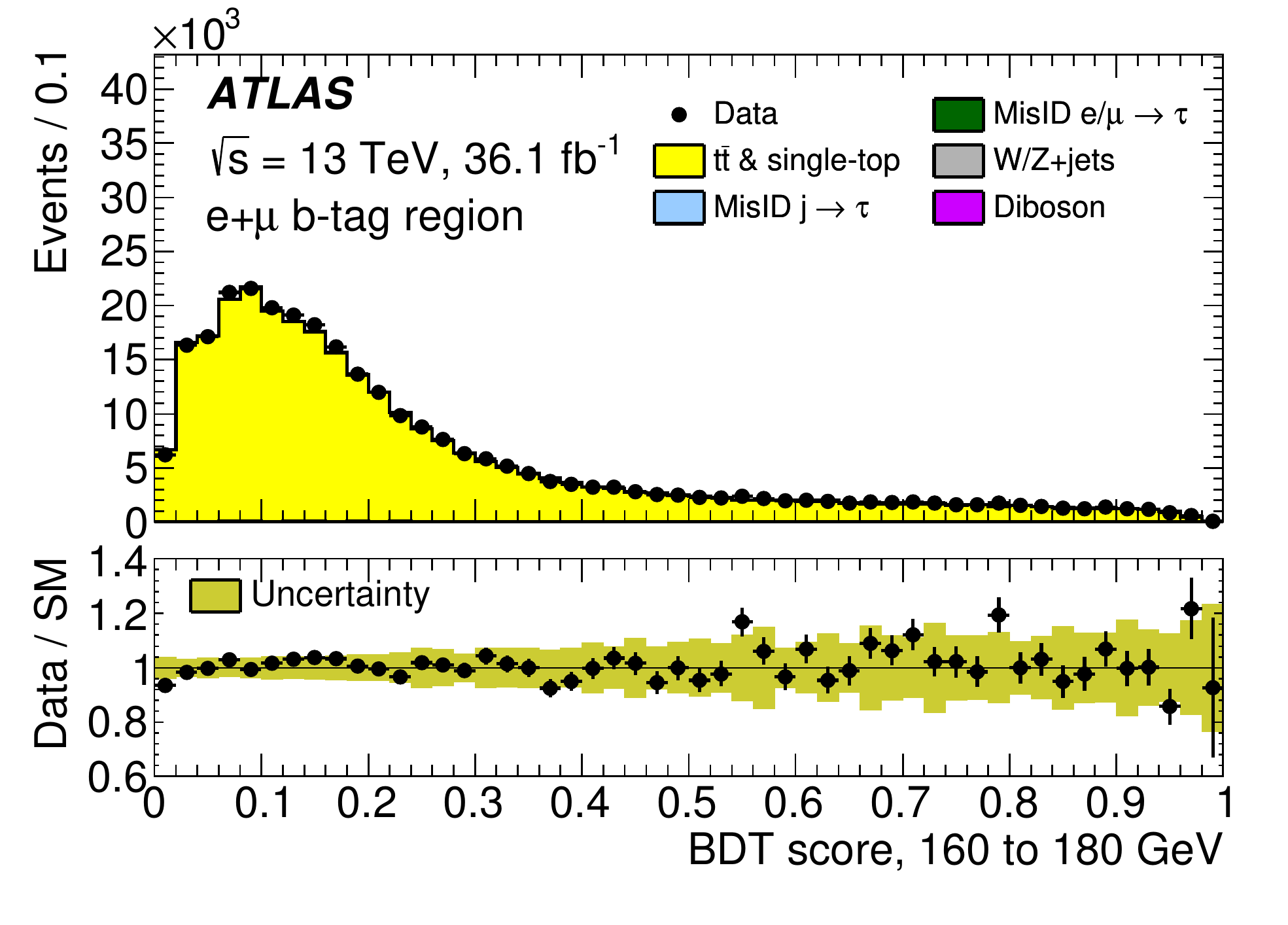}}
\subfloat[\label{fig:taulepBDTscoreDLT:d}]{\includegraphics[width=0.45\linewidth]{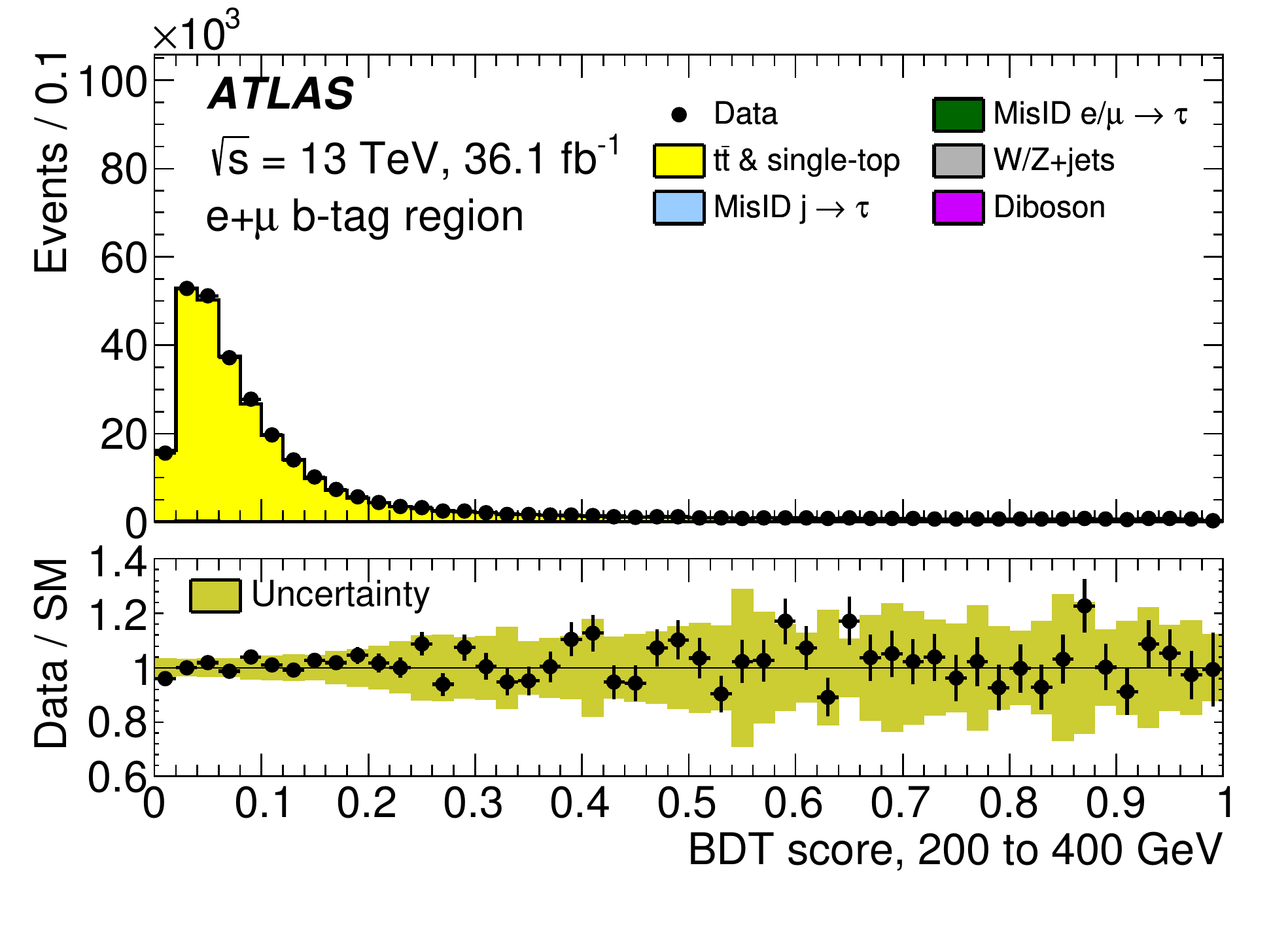}} \\
\subfloat[\label{fig:taulepBDTscoreDLT:e}]{\includegraphics[width=0.45\linewidth]{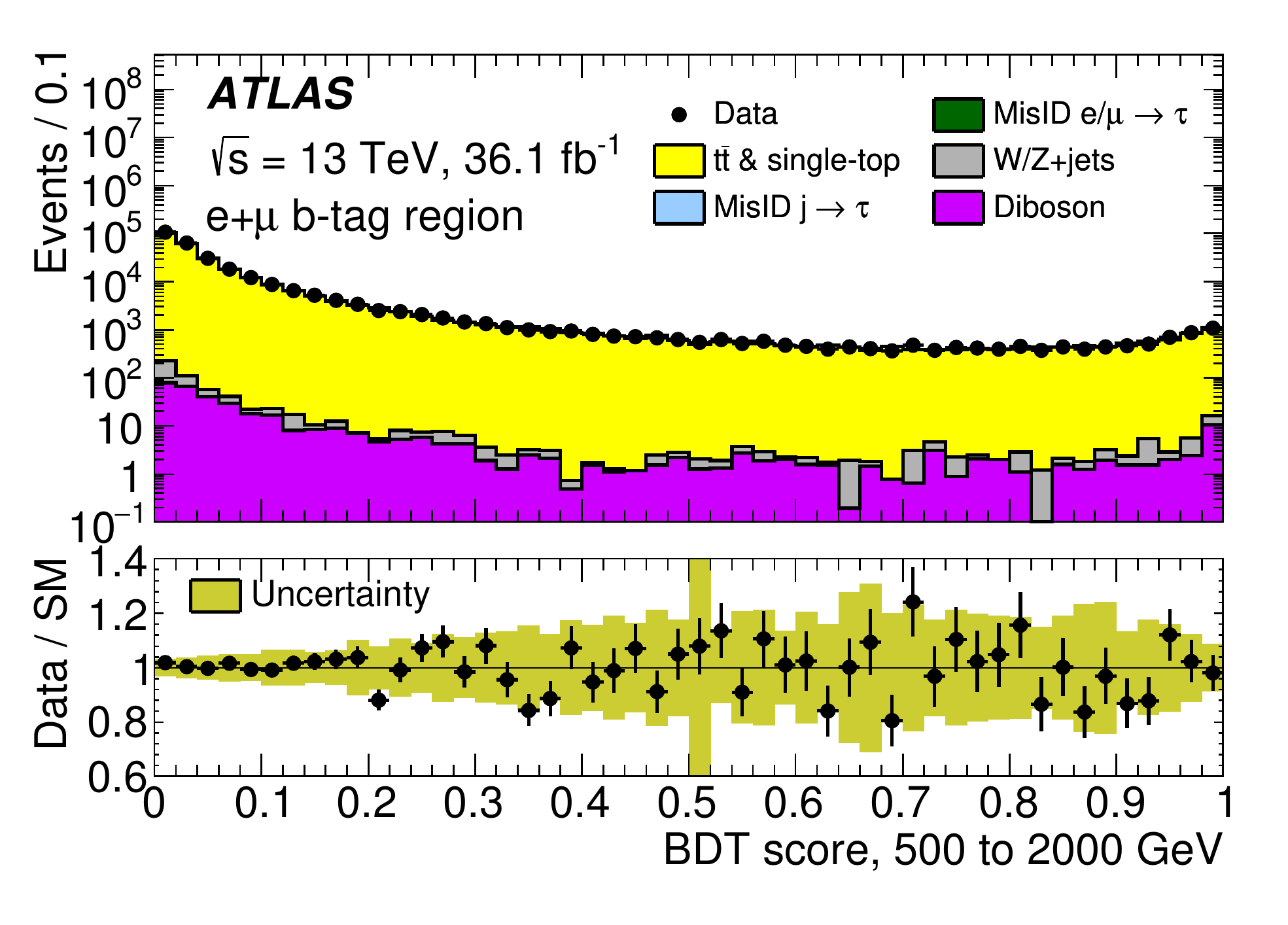}}
\caption{\label{fig:taulepBDTscoreDLT:main} BDT score distribution for
the predicted backgrounds and data in a region with the same event selection as for 
the $\tauhadvis$+lepton channel, except for the requirement of an $e\mu$ pair instead 
of the $e/\mu + \tauhadvis$ pair (as described in the text). The five $H^{+}$ mass range 
trainings are shown. The lower panel of each plot shows the ratio of data to the SM background 
prediction.  The uncertainty bands include all statistical and systematic uncertainties. 
}
\end{figure}

\begin{figure}[htbp]
  \centering
\subfloat[\label{fig:taulepBDTscoreWZ:a}]{\includegraphics[width=0.45\linewidth]{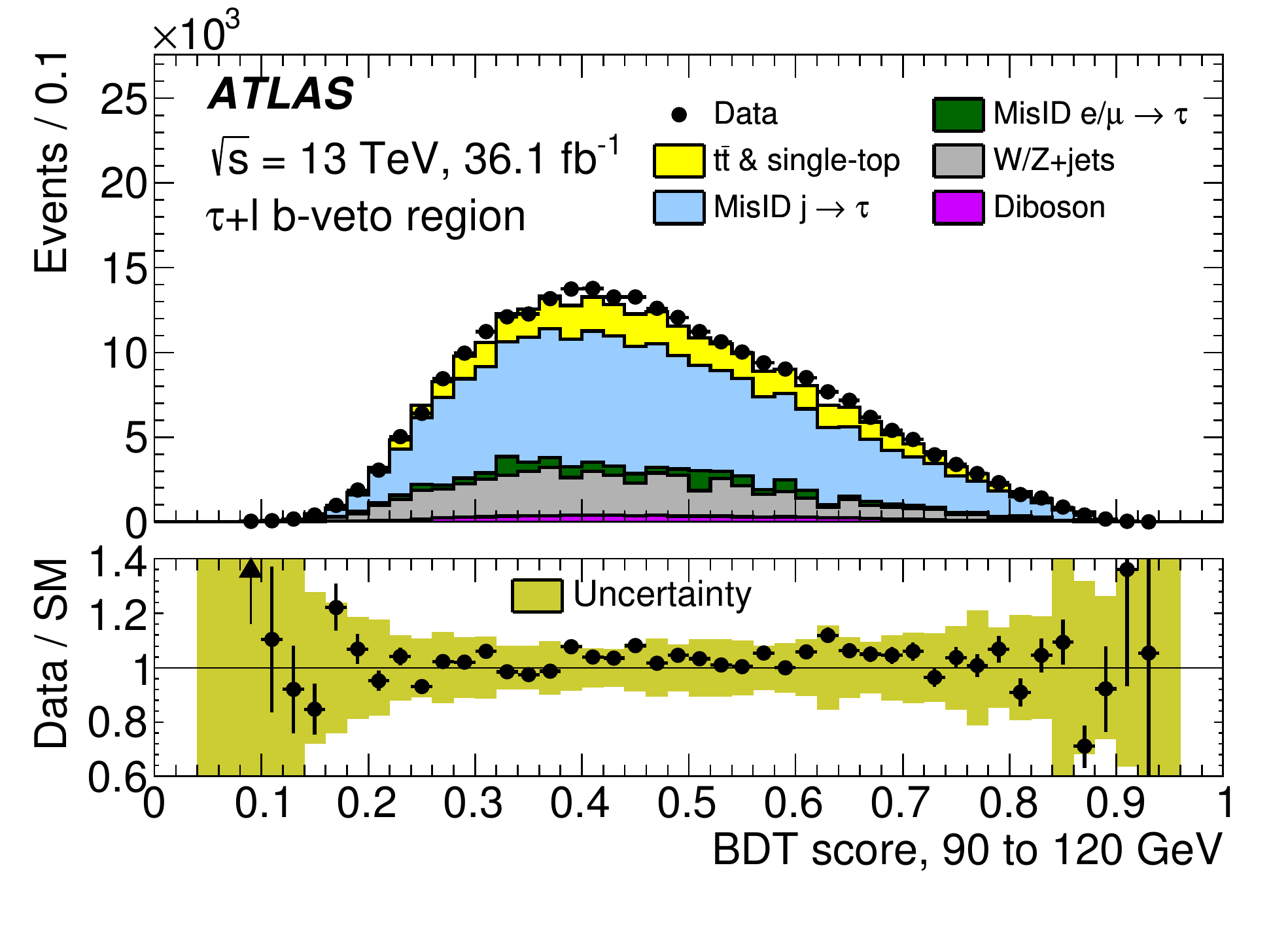}}
\subfloat[\label{fig:taulepBDTscoreWZ:b}]{\includegraphics[width=0.45\linewidth]{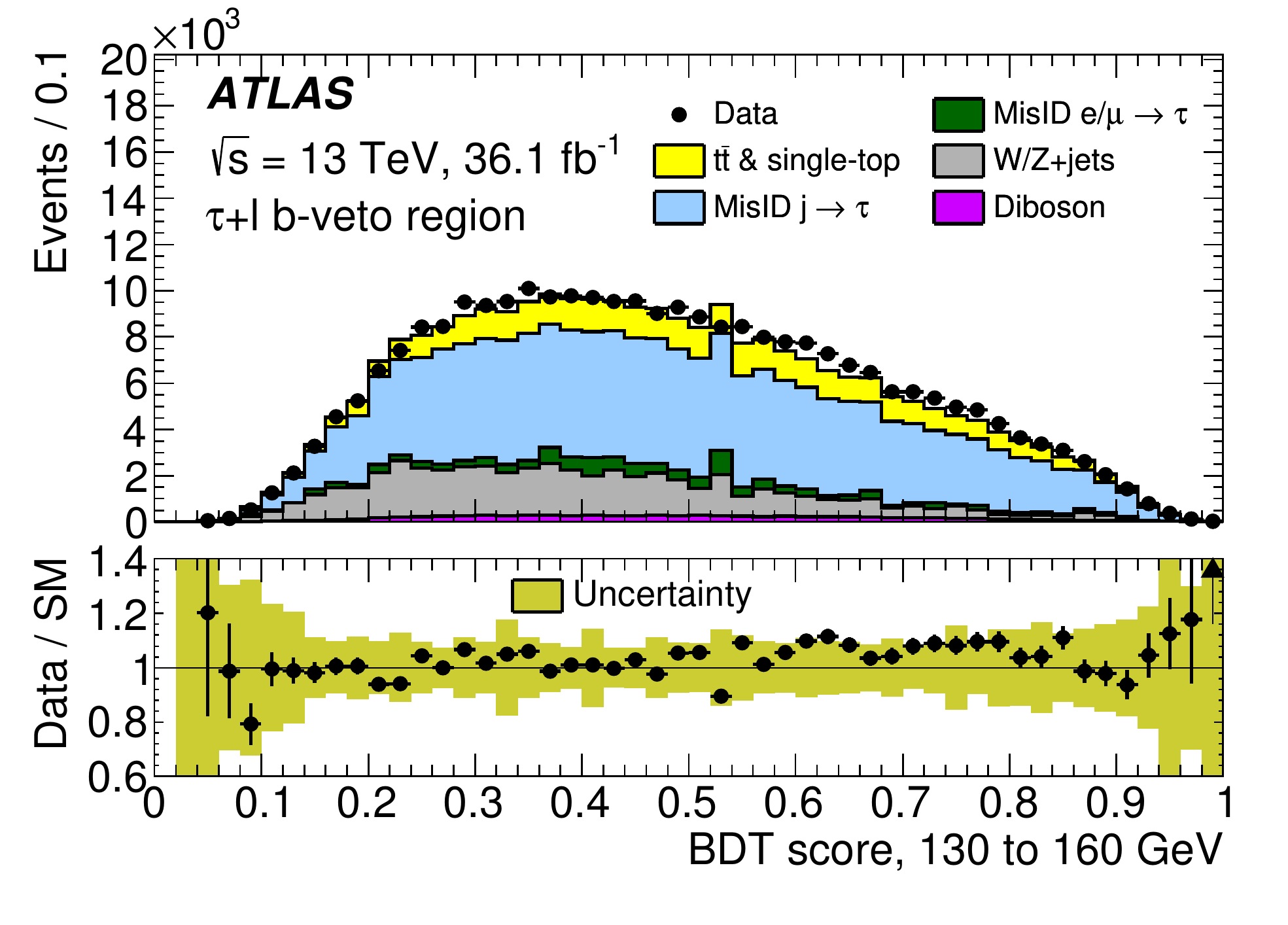}} \\
\subfloat[\label{fig:taulepBDTscoreWZ:c}]{\includegraphics[width=0.45\linewidth]{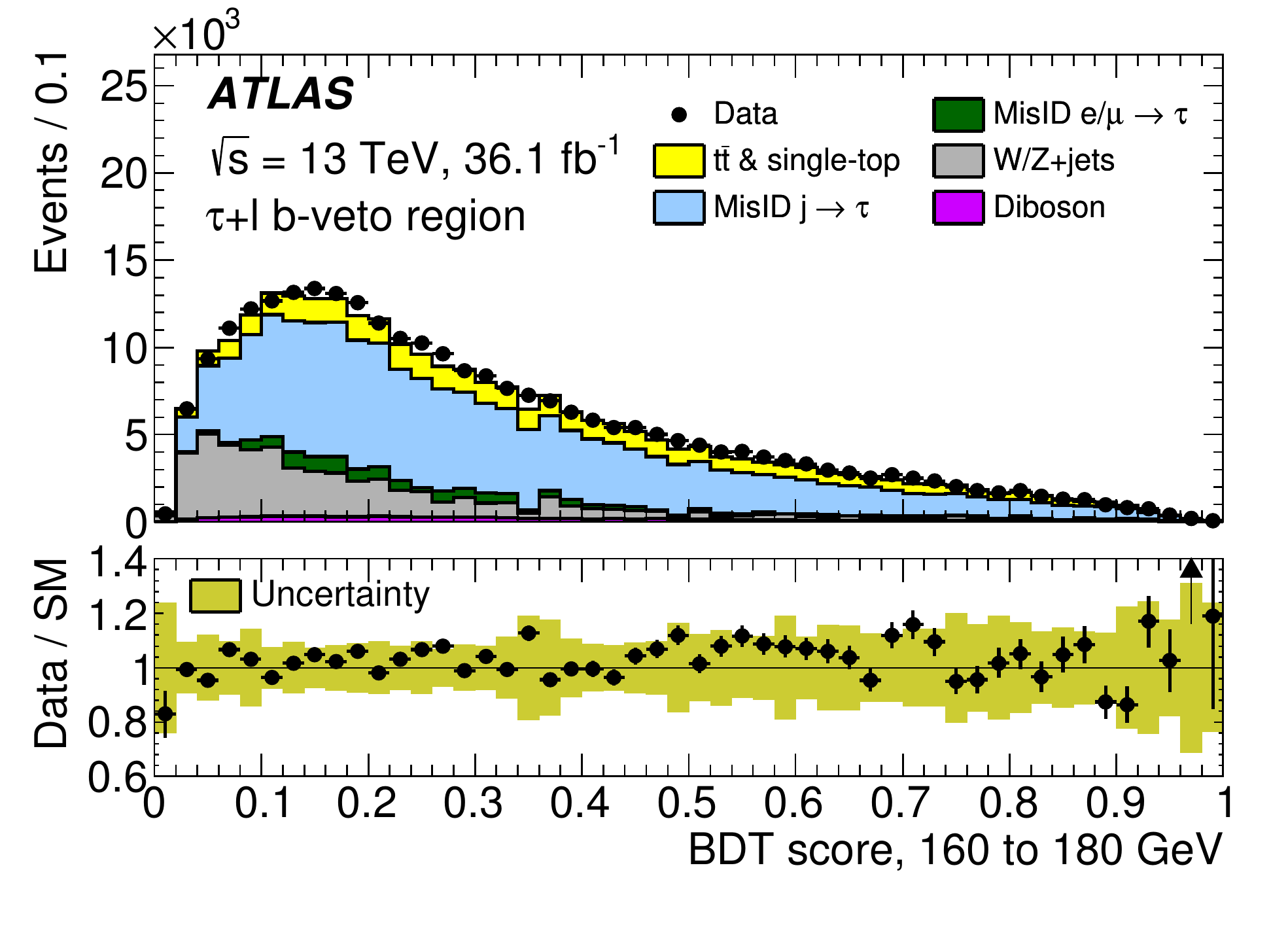}}
\subfloat[\label{fig:taulepBDTscoreWZ:d}]{\includegraphics[width=0.45\linewidth]{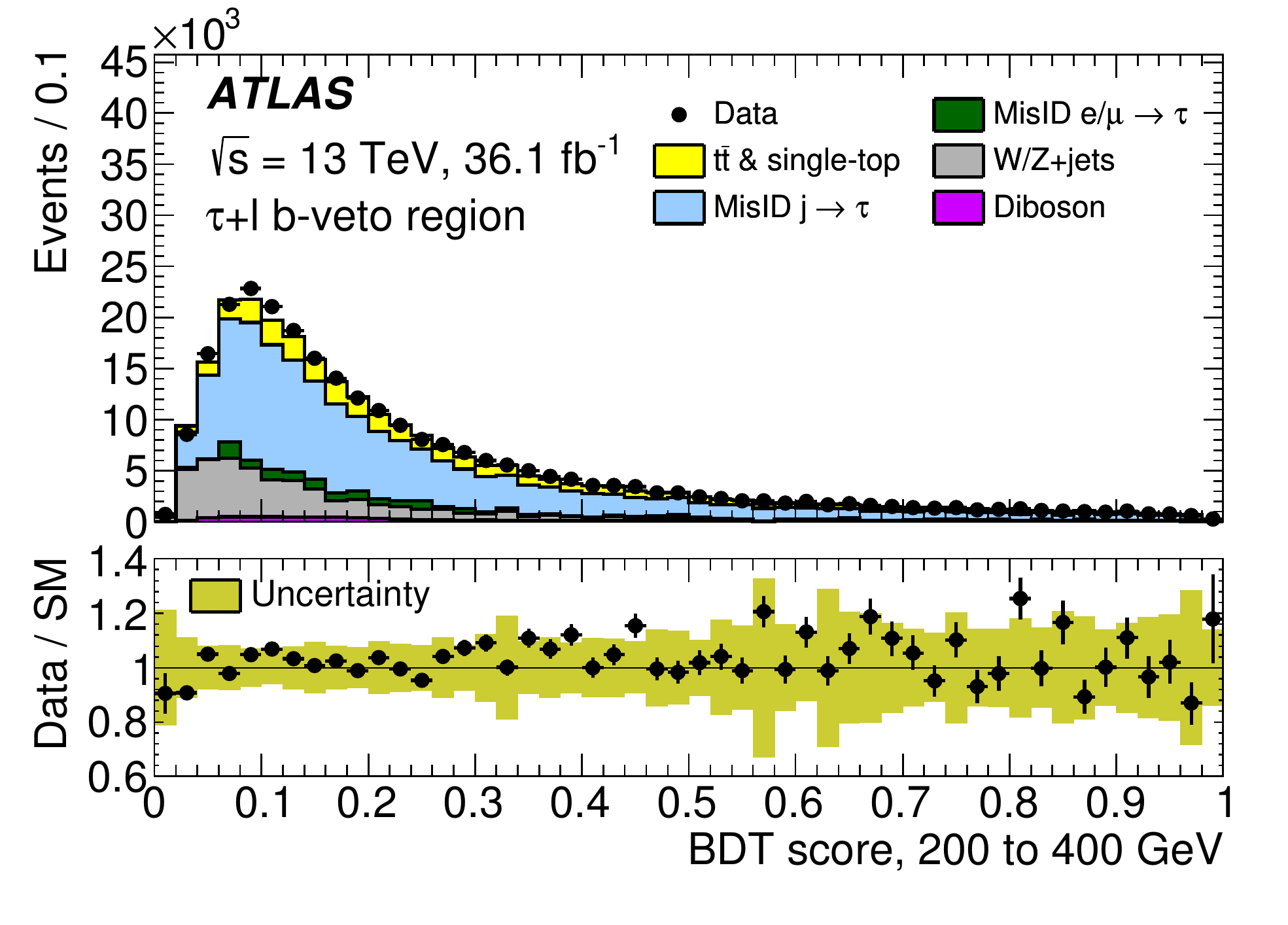}} \\
\subfloat[\label{fig:taulepBDTscoreWZ:e}]{\includegraphics[width=0.45\linewidth]{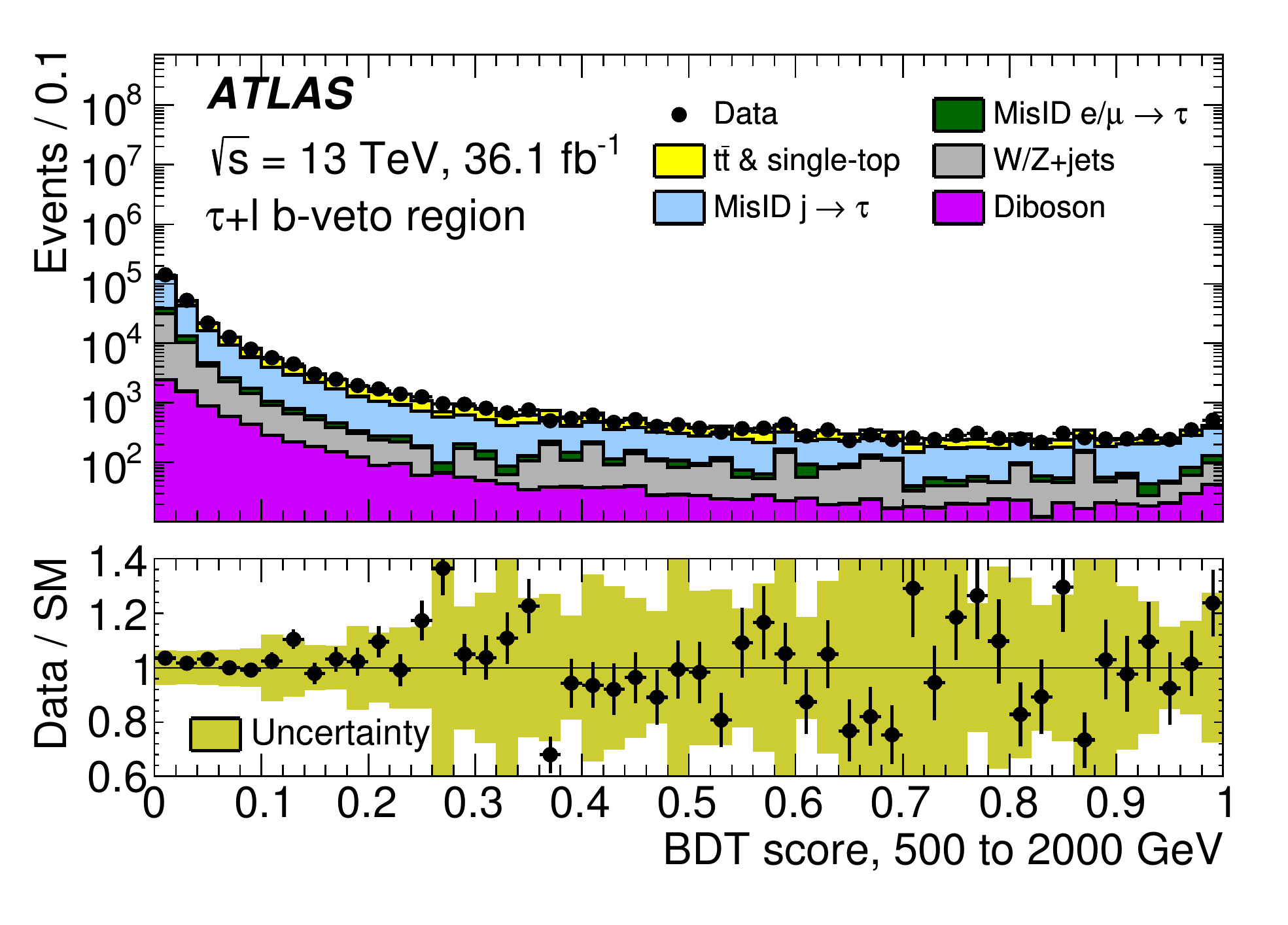}}
\caption{\label{fig:taulepBDTscoreWZ:main} BDT score distribution for
the predicted backgrounds and data in a region with the same event selection as for the 
$\tauhadvis$+lepton channel, except that it has exactly zero $b$-tagged jet
(as described in the text). The five $H^{+}$ mass range trainings are shown. 
The lower panel of each plot shows the ratio of data to the SM background prediction.  The uncertainty bands include all statistical and systematic uncertainties. 
}
\end{figure}

The modelling of the backgrounds, especially \ttbar\ and events with a misidentified \tauhadvis\ candidate, 
is validated in signal-depleted regions. A region enriched in \ttbar\ events is defined with the same 
event selection as the $\tauhadvis$+lepton channel, but with the requirement of having an $e\mu$ 
pair (with $\pt$ above 30~\GeV\ for both the electron and muon) instead of the $e/\mu + \tauhadvis$ pair. 
This control region is included as a single-bin distribution in the statistical analysis described in 
Section~\ref{sec:results}. The modelling of the background with misidentified \tauhadvis\ objects is 
validated in a region that is defined with the same selection criteria as the $\tauhadvis$+lepton 
channel, but with a veto on $b$-tagged jets. Predicted and measured BDT score distributions are compared 
in the two regions discussed above, and they are found to be in good agreement prior to any fit to the data, 
as shown in Figures~\ref{fig:taulepBDTscoreDLT:main} and~\ref{fig:taulepBDTscoreWZ:main}. 
This procedure validates the modelling of the two main SM background processes.

\section{Systematic uncertainties}
\label{sec:systematic_uncertainties}
Several sources of systematic uncertainty affect the normalisation of the signal
and background processes, as well as the shape of the BDT score distribution
used as the final discriminant. Individual sources of systematic uncertainty
are assumed to be uncorrelated. However, when the systematic variations are 
applied to different samples of simulated events, correlations of a given 
systematic uncertainty are taken into account across processes. 

All instrumental sources of systematic uncertainty, i.e.\ the reconstruction and
identification efficiencies, as well as the energy scales and resolutions of electrons, 
muons, ($b$-tagged) jets and \tauhadvis\ candidates, are considered, including their
impact on the reconstructed \MET. In both the $\tau_{\text{had-vis}}$+jets and
$\tau_{\text{had-vis}}$+lepton channels, the dominant systematic uncertainties
come from the jet energy scale~\cite{Aaboud:2017jcu} (between 1\% and 4.5\% depending on the jet 
$\pt$), the $b$-tagging efficiency~\cite{btagunc} (ranging from 2\% to 10\% depending on 
the jet $\pt$), the reconstruction and identification efficiencies of \tauhadvis\ 
candidates (3\% and 6\%, respectively), and their energy scale (2--3\%)~\cite{ATLAS-CONF-2017-029}. 
Additional uncertainties based on multi-jet data and single-particle response 
are added for the identification efficiency and the energy scale of \tauhadvis\ 
objects at high \pt, respectively. The probability for electrons to be misidentified 
as \tauhadvis\ is measured with a precision of 3--14\%. The uncertainty of 2.1\% in 
the integrated luminosity is applied directly to the event yields of all simulated events. 
In the $\tau_{\text{had-vis}}$+lepton channel, the impact of the systematic uncertainty 
in the single-lepton trigger efficiency is at most 1\%. In the $\tau_{\text{had-vis}}$+jets 
channel, the efficiency of the \MET\ trigger is measured in a control region of the data, 
as described in Ref.~\cite{Aaboud:2016dig}. The associated systematic uncertainty in 
the event yield of the signal region is 1.4\%.

In the estimation of backgrounds with jets misidentified as \tauhadvis\
candidates, the sources of systematic uncertainty are:
\begin{itemize}
\item the loose requirement on the \tauhadvis\ BDT output score used in the
definition of the anti-\tauhadvis\ sample, which modifies the
corresponding fractions of quark- and gluon-initiated jets, as well as the
event topology (assessed by considering the shape of the final discriminant
obtained for two alternative thresholds for the BDT output score that are symmetric
around the nominal threshold);
\item the level of contamination of true \tauhadvis\ candidates fulfilling the anti-\tauhadvis\
selection (varied by 50\%);
\item the statistical uncertainties in the event yields entering the computation of FFs, in each 
bin of their parameterisation and for each control region;
\item the statistical uncertainty of the best-fit value of $\alpha_{\text{MJ}}$; 
\item the modelling of heavy-flavour jets mimicking \tauhadvis\ candidates, obtained by
computing the fake factors separately for light- and heavy-quark-initiated jets, as in
Ref.~\cite{Aaboud:2016dig}, and comparing those with the nominal predictions, then using
the difference as a systematic uncertainty;
\item for the $\Upsilon$ distribution only, the uncertainty in the inverse transform sampling method, 
taken as the difference between the computations obtained in the two control regions where the FFs are 
measured.
\end{itemize}

The dominant background with a true \tauhadvis\ candidate is from \ttbar\ pairs
and single-top-quark events. 
A normalisation factor is computed for this background by including the control region 
of the \tauhadvis+lepton channel with an $e\mu$ pair and at least one $b$-jet as a single-bin 
distribution in the statistical analysis. Systematic uncertainties in the modelling of 
the \ttbar\ background are included. Those due to the choice of parton shower 
and hadronisation models are derived by comparing \ttbar\ events generated with 
{\sc Powheg-Box v2} interfaced to either {\sc Pythia6} or \Herwigpp~\cite{Bahr:2008pv}. 
The systematic uncertainties arising from additional radiation, which modify the jet production 
rate, are computed with the same packages as for the baseline \ttbar\ event generation, by varying 
the shower radiation, the factorisation and renormalisation scales, as well as the NLO radiation. 
The uncertainty due to the choice of matrix-element generator is evaluated
by comparing \ttbar\ samples generated with {\sc MadGraph5\_aMC@NLO}
or {\sc Powheg-Box}, both using the CT10 PDF set and interfaced to
\Herwigpp. The impacts of the three systematic uncertainties listed above 
on the event yield of the \ttbar\ background are, respectively, 14\%, 4\%, 13\% 
in the \tauhadvis+jets channel and 13\%, 8\%, 9\% in the \tauhadvis+lepton channel. 
For the $W$+jets and $Z$+jets backgrounds, uncertainties of 35\% and 40\% based on 
variations of the scales in {\sc Sherpa} are considered, respectively. An additional 
uncertainty in the heavy-flavour jet modelling is derived by comparing the predictions 
of {\sc Sherpa} and {\sc MadGraph5\_aMC@NLO} interfaced to {\sc Pythia6}: it is about 
6\% in the \tauhadvis+jets channel and 14\% in the \tauhadvis+lepton channel. For 
the small diboson backgrounds, an uncertainty of 50\%, arising from the inclusive
cross-section and additional jet production, is used~\cite{ATL-PHYS-PUB-2016-002}.

Systematic uncertainties in the \Hp\ signal generation are estimated
as follows. The uncertainty arising from the missing higher-order corrections is 
assessed by varying the factorisation and renormalisation scales up and down by a 
factor of two. The largest variation of the signal acceptance is then symmetrised 
and taken as the scale uncertainty, 4--8\% depending on the \Hp\ mass hypothesis. 
The signal acceptances are computed with various PDF sets, following the PDF4LHC 
prescriptions~\cite{Butterworth:2015oua}, and their envelope is taken as a systematic 
uncertainty. The impact of A14 tune variations on the signal acceptance
is estimated by adding in quadrature the excursions from a subset of tune
variations that cover underlying-event and jet-structure effects, as well as
different aspects of extra jet production. This uncertainty amounts to 8--10\%.

In the low- and intermediate-mass $\Hp$ search, the main systematic uncertainties arise 
from the estimation of the background with misidentified \tauhadvis\ candidates, as well 
as the reconstruction and identification of \tauhadvis\ candidates. For large $\Hp$ masses, 
systematic uncertainties from the signal modelling and the estimation of the background with 
misidentified \tauhadvis\ candidates dominate, but the search is also limited by the number 
of selected events. The impact of the systematic uncertainties on the sensitivity of the analysis 
is discussed in Section~\ref{sec:results}.

\section{Results}
\label{sec:results}
The statistical interpretation is based on a simultaneous fit of the
parameter of interest, e.g.\ $\mu \equiv \sigma(pp \to tbH^+) \times 
\mathrm{\cal{B}}(H^+ \to \tau\nu)$, and the nuisance parameters $\mathbf{\theta}$ that encode
statistical and systematic uncertainties, by means of a negative log-likelihood minimisation.
The test statistic $\tilde{q}_{\mu}$~\cite{Cowan:2010js} used to test the compatibility of the 
data with the background-only and signal+background hypotheses is computed from the profile 
likelihood ratio, and the asymptotic approximation is used throughout the statistical analysis.
Three signal regions and one control region enriched in \ttbar\ events are considered
in the simultaneous fit:
\begin{itemize}
\item Binned likelihood functions are used for the BDT score distributions in the three signal 
regions ($\tauhadvis$+jets, $\tauhadvis$+electron and $\tauhadvis$+muon). The binning 
of the discriminating variable is optimised to maximise the sensitivity of the analysis prior to looking 
at the data in the signal regions.
\item A single-bin likelihood is used in the control region enriched in  \ttbar\ events, defined with
the same event selection as the $\tauhadvis$+lepton channel, but with the requirement of
an $e\mu$ pair instead of the $e/\mu + \tauhadvis$ pair.
\end{itemize}

The expected number of events for all SM processes and the measured event yields 
in the signal regions are shown in Tables~\ref{tab:event_yields_SR_taujets}
and~\ref{tab:event_yields_SR_taulep}, prior to using the multivariate discriminant 
and applying the statistical fitting procedure. The contributions from hypothetical 
charged Higgs bosons are also shown, assuming a mass of 170~\GeV\ or 1000~\GeV, and with 
$\sigma(pp \to tbH^+) \times \mathrm{\cal{B}}(H^+ \to \tau\nu)$ set to the prediction 
from the hMSSM scenario for $\tan\beta= 40$, as computed using Refs.~\cite{Degrande:2016hyf} 
and~\cite{Heinemeyer:2013tqa,Degrande:2015vpa,Flechl:2014wfa,Dittmaier:2009np,Berger:2003sm}
for the production cross-section and {\sc HDECAY}~\cite{Djouadi:1997yw} for the branching 
fraction. The signal acceptances for a charged Higgs boson mass hypothesis of 170~\GeV, as 
evaluated in a sample of simulated events where both the $\tau$-lepton and the top-quark decay 
inclusively, are 0.9\%, 0.6\% and 0.5\% in the $\tauhadvis$+jets, $\tauhadvis$+electron and 
$\tauhadvis$+muon signal regions, respectively. They become 11.6\%, 0.9\% and 1.2\% 
for a charged Higgs boson mass of 1~\TeV. The event yields observed in \LUMI\ of 
data collected at 13~\TeV\ are consistent with the expected SM backgrounds, but 
very little sensitivity to $\Hp$ signals can be obtained from the comparison of 
event yields only.

\begin{figure}
  \centering
\subfloat[\label{fig:taujets_SR_BDTscores_body_a}]{\includegraphics[width=0.45\linewidth]{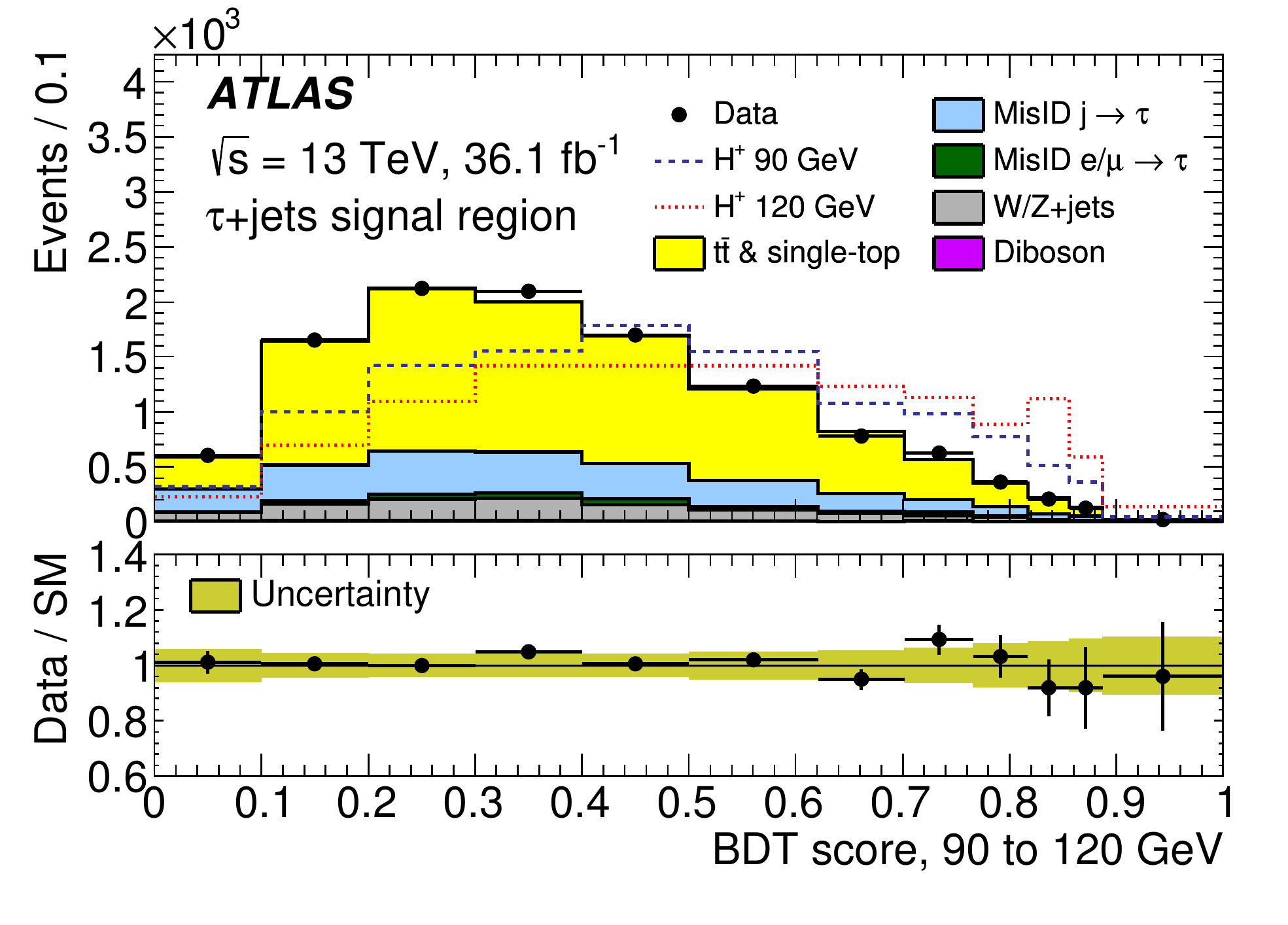}}
\subfloat[\label{fig:taujets_SR_BDTscores_body_b}]{\includegraphics[width=0.45\linewidth]{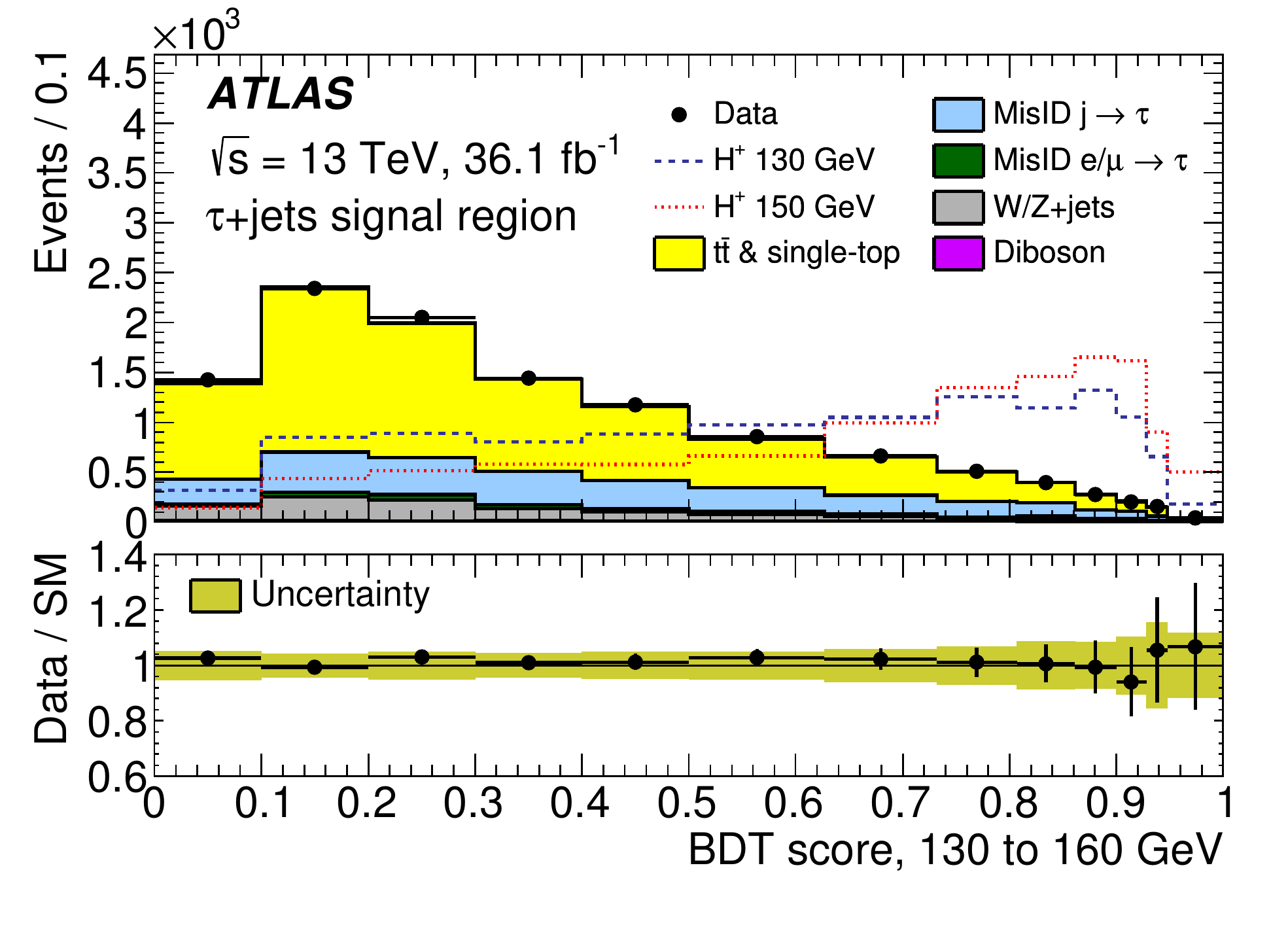}} \\
\subfloat[\label{fig:taujets_SR_BDTscores_body_c}]{\includegraphics[width=0.45\linewidth]{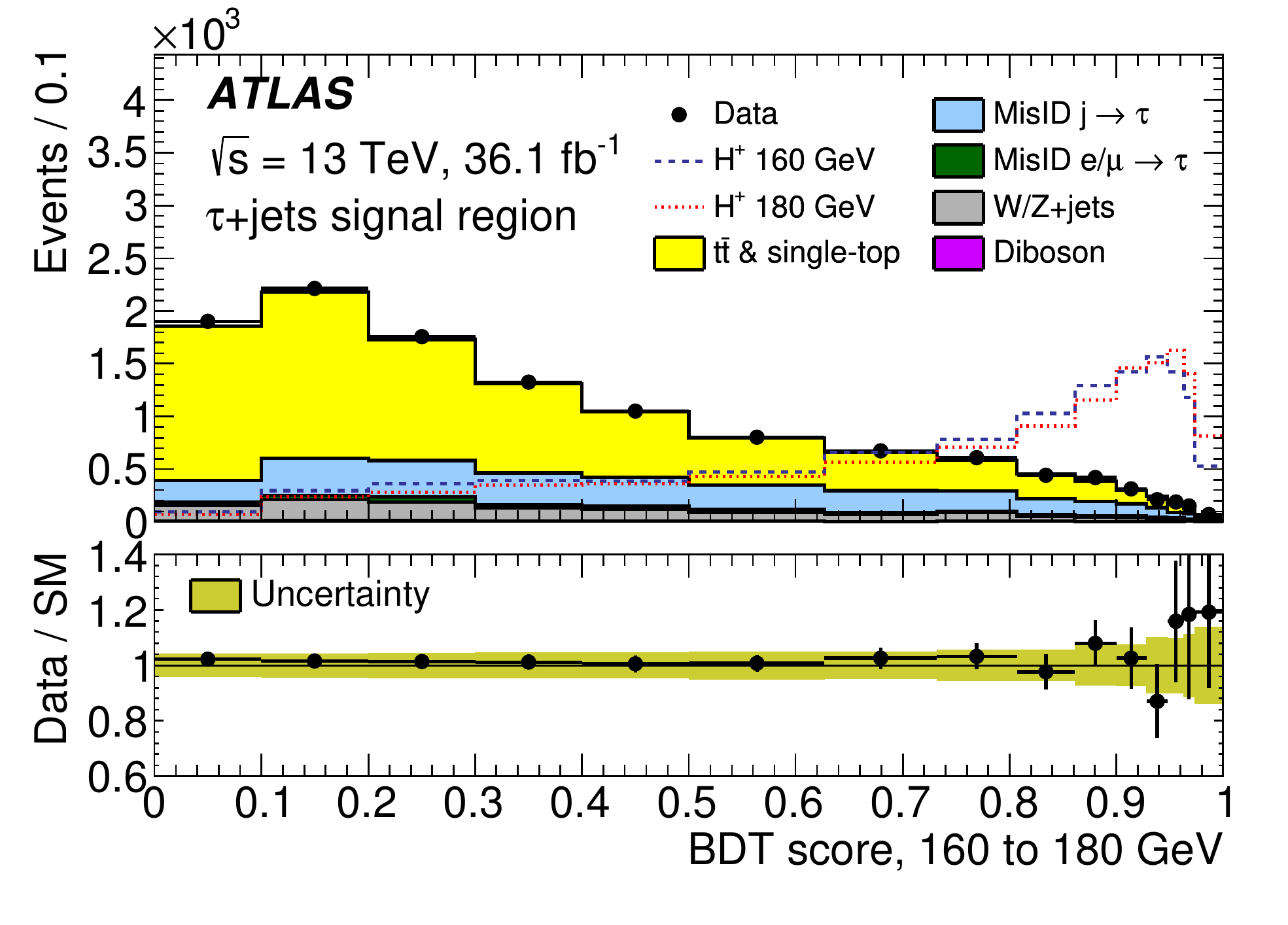}}
\subfloat[\label{fig:taujets_SR_BDTscores_body_d}]{\includegraphics[width=0.45\linewidth]{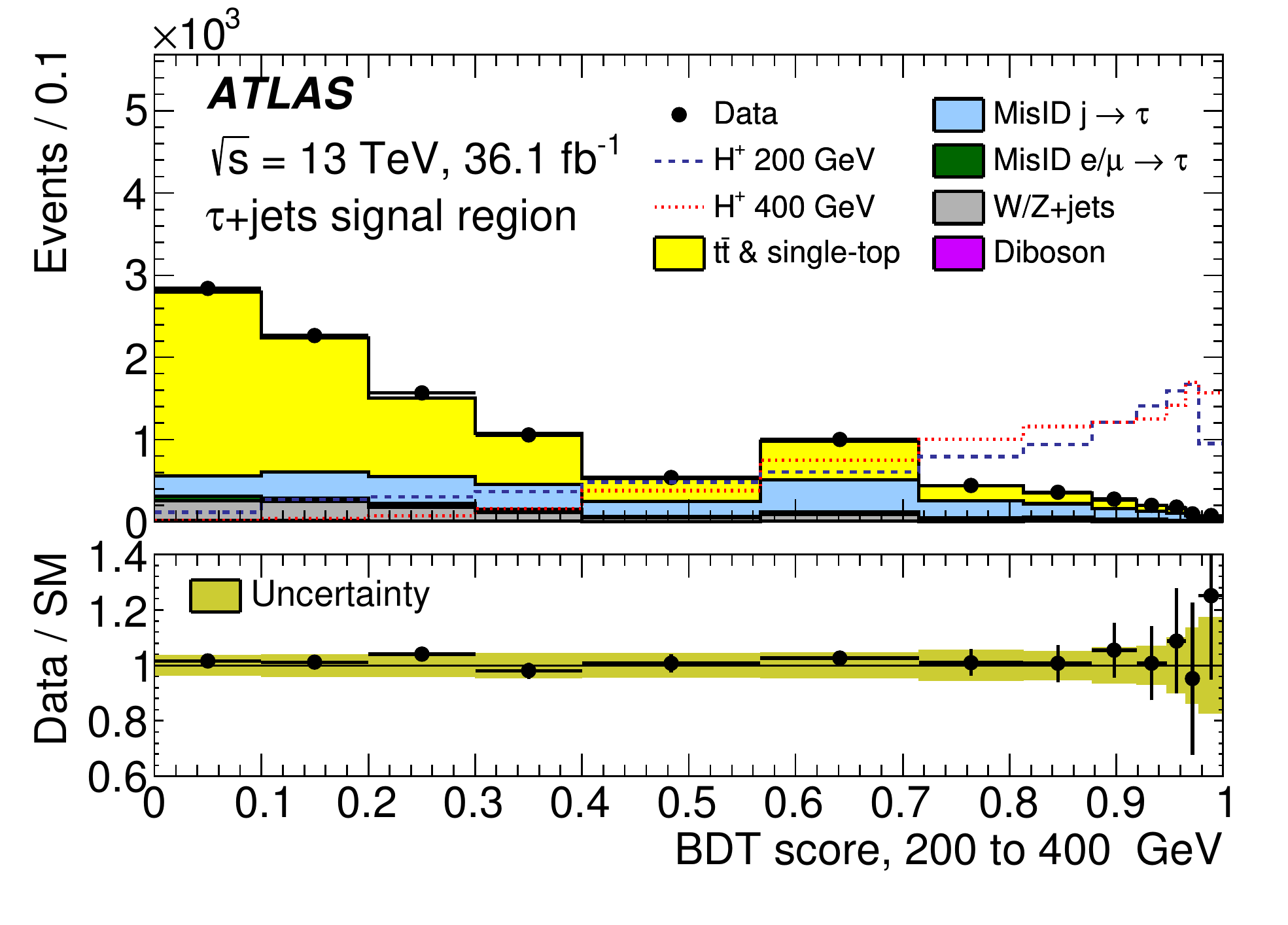}} \\
\subfloat[\label{fig:taujets_SR_BDTscores_body_e}]{\includegraphics[width=0.45\linewidth]{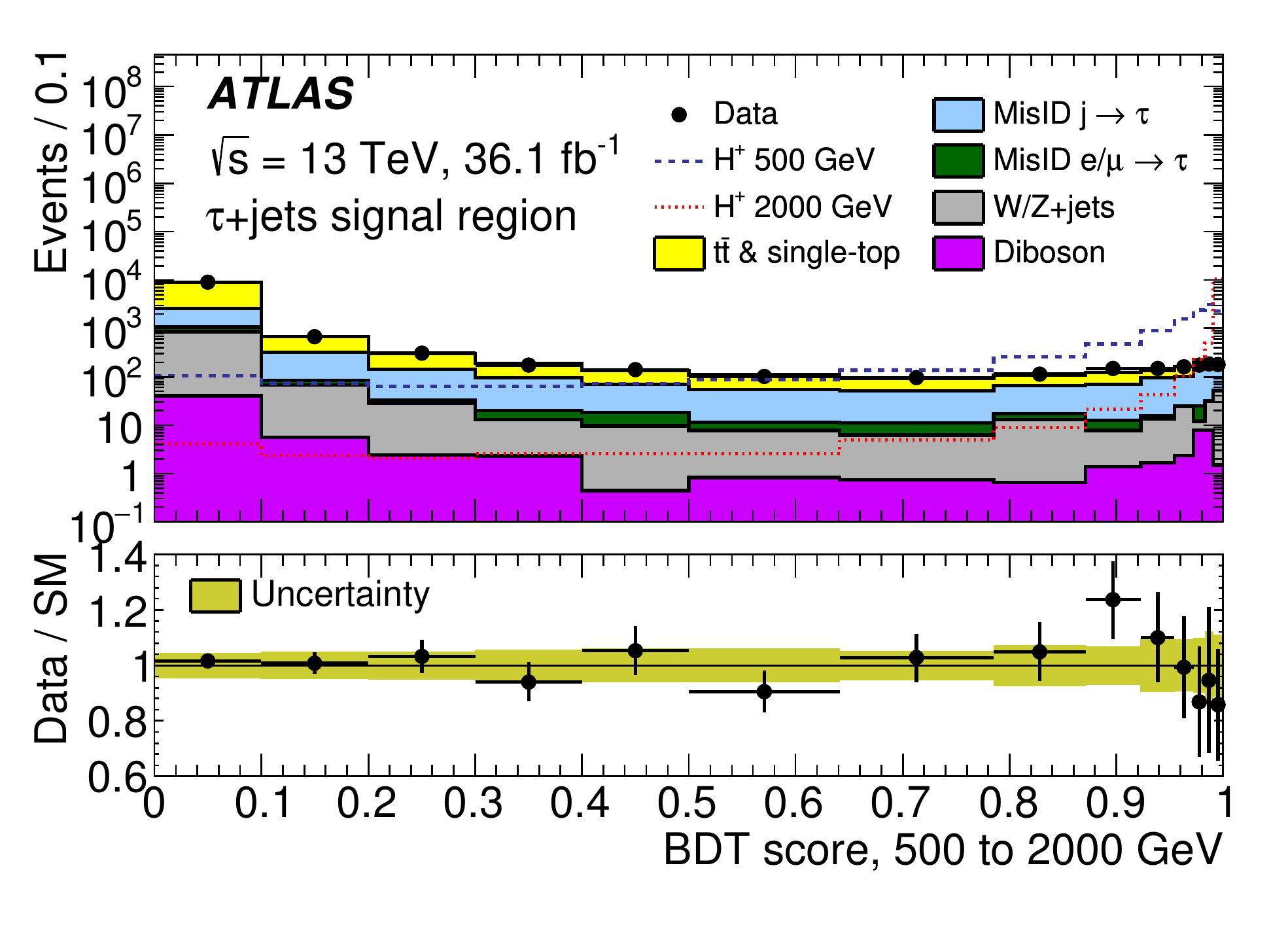}}
  \caption{\label{fig:taujets_SR_BDTscores_body} BDT score distributions in the
signal region of the $\tauhadvis$+jets channel, in the five mass ranges used for the
BDT trainings, after a fit to the data with the background-only hypothesis. 
The lower panel of each plot shows the ratio of data to the SM background prediction.  The uncertainty bands include all statistical and systematic uncertainties. 
The normalisation of the signal (shown for illustration) corresponds to the integral of the background.}
\end{figure}

\begin{figure}
\centering
\subfloat[\label{fig:taulepBDTscoreSR1_body_a}]{\includegraphics[width=0.45\linewidth]{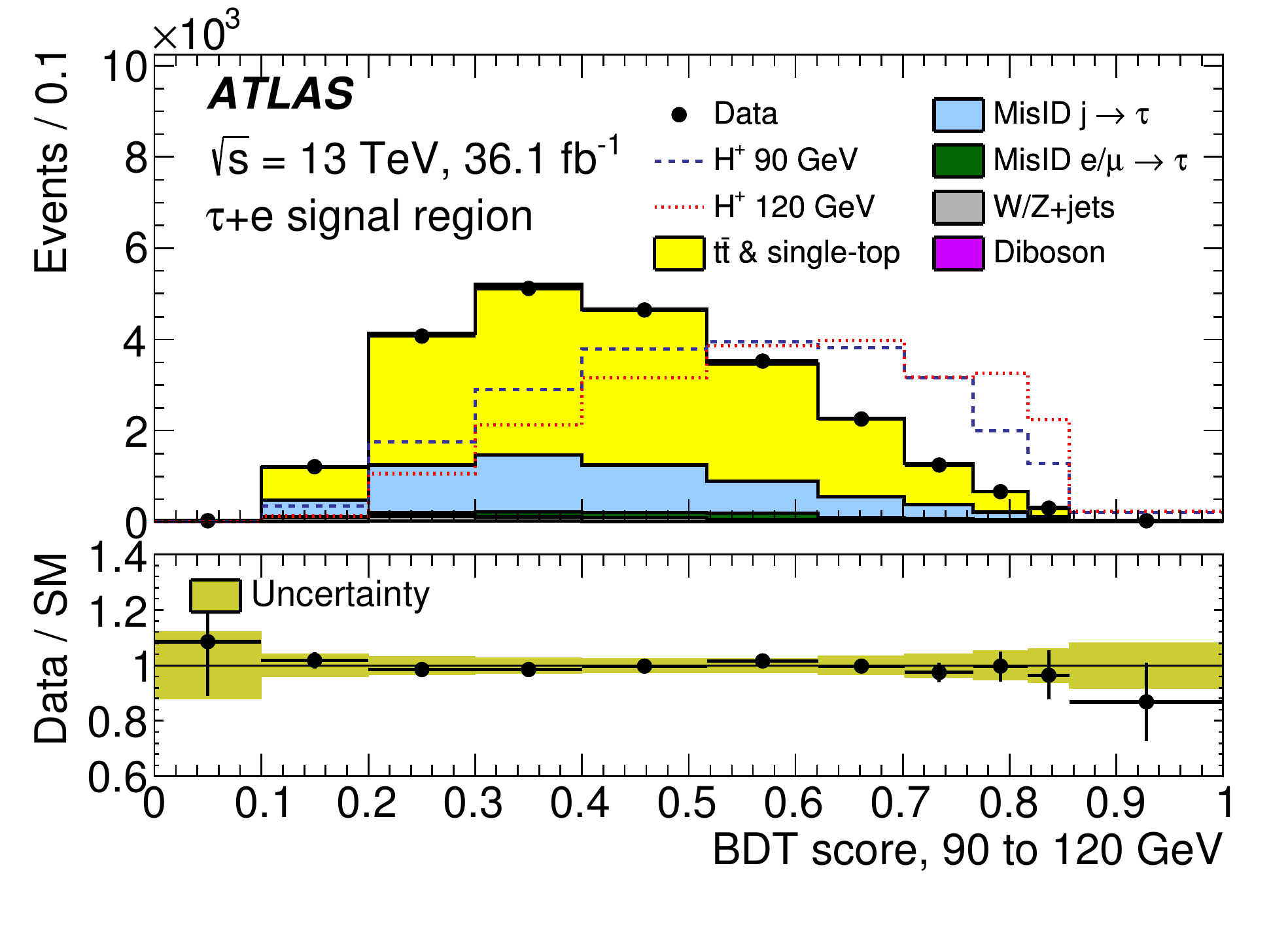}}
\subfloat[\label{fig:taulepBDTscoreSR1_body_b}]{\includegraphics[width=0.45\linewidth]{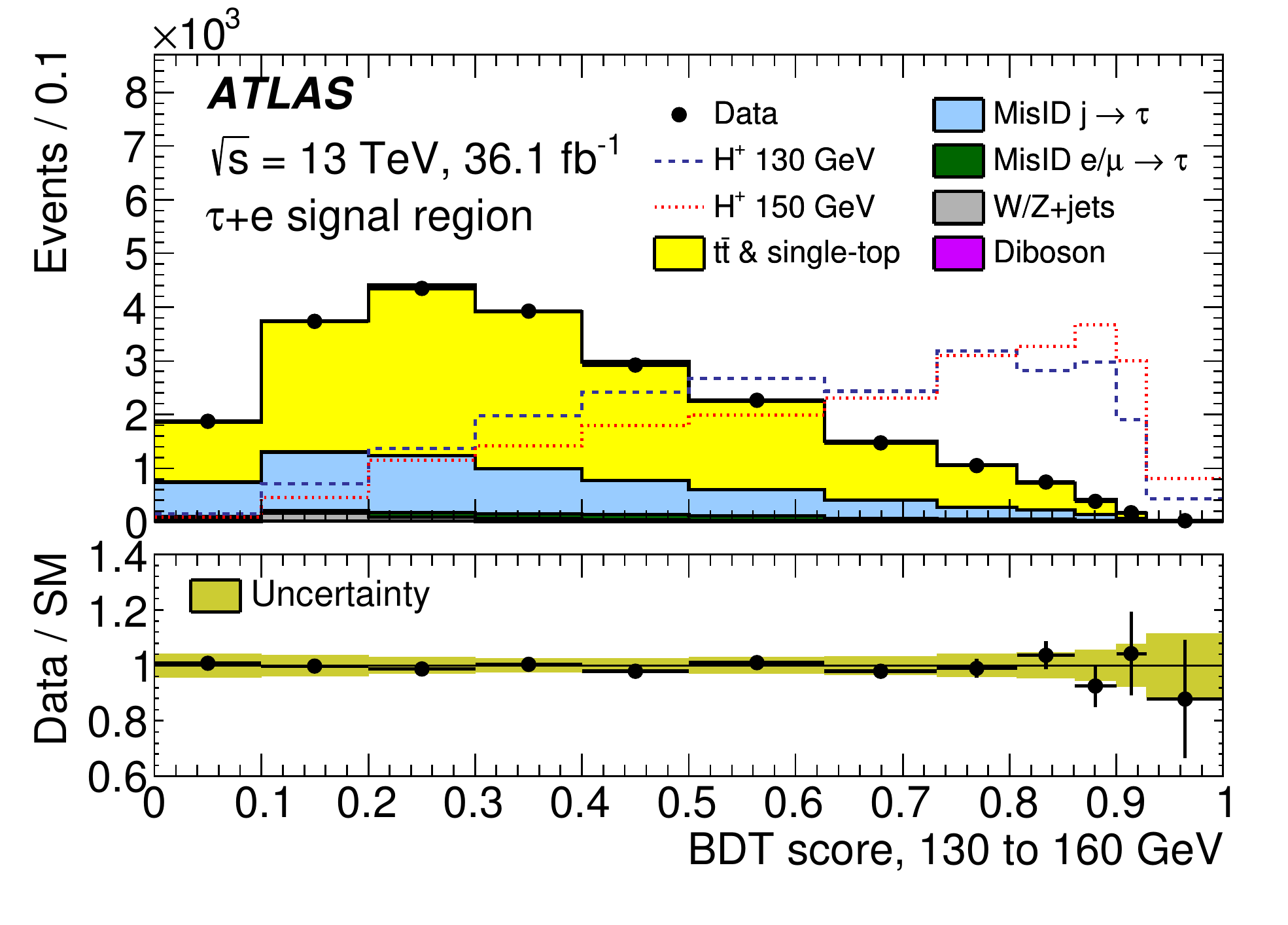}} \\
\subfloat[\label{fig:taulepBDTscoreSR1_body_c}]{\includegraphics[width=0.45\linewidth]{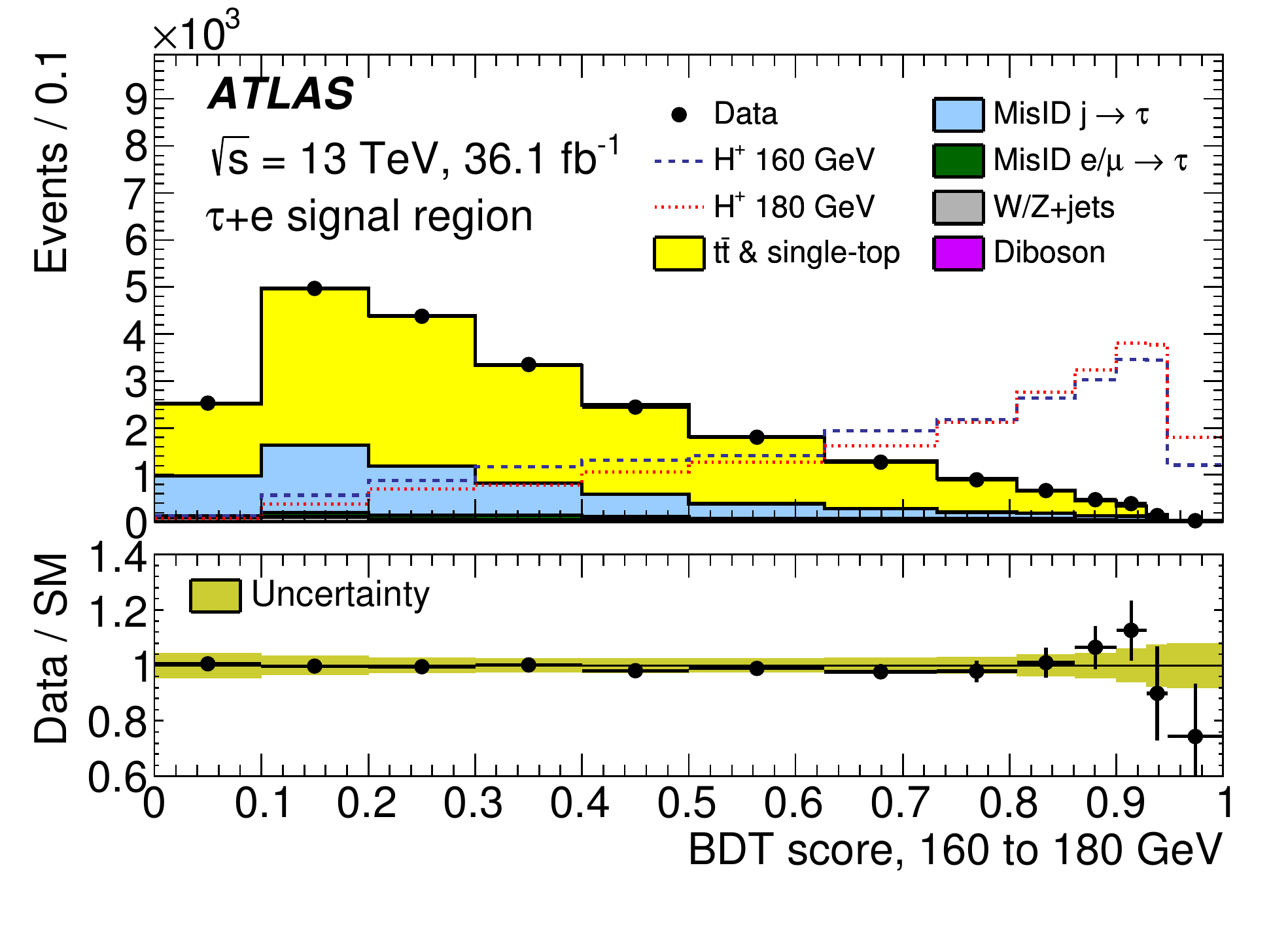}}
\subfloat[\label{fig:taulepBDTscoreSR1_body_d}]{\includegraphics[width=0.45\linewidth]{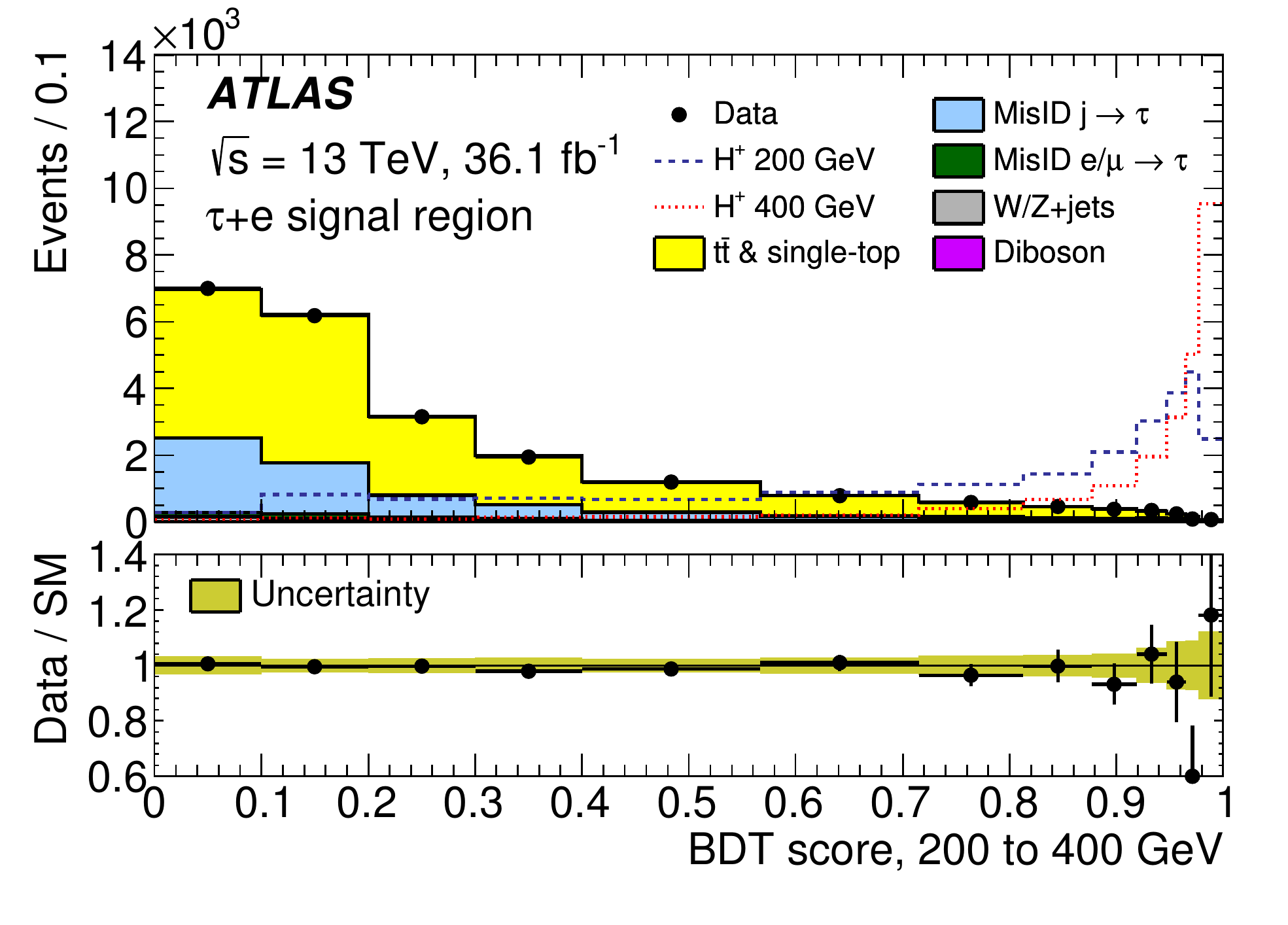}} \\
\subfloat[\label{fig:taulepBDTscoreSR1_body_e}]{\includegraphics[width=0.45\linewidth]{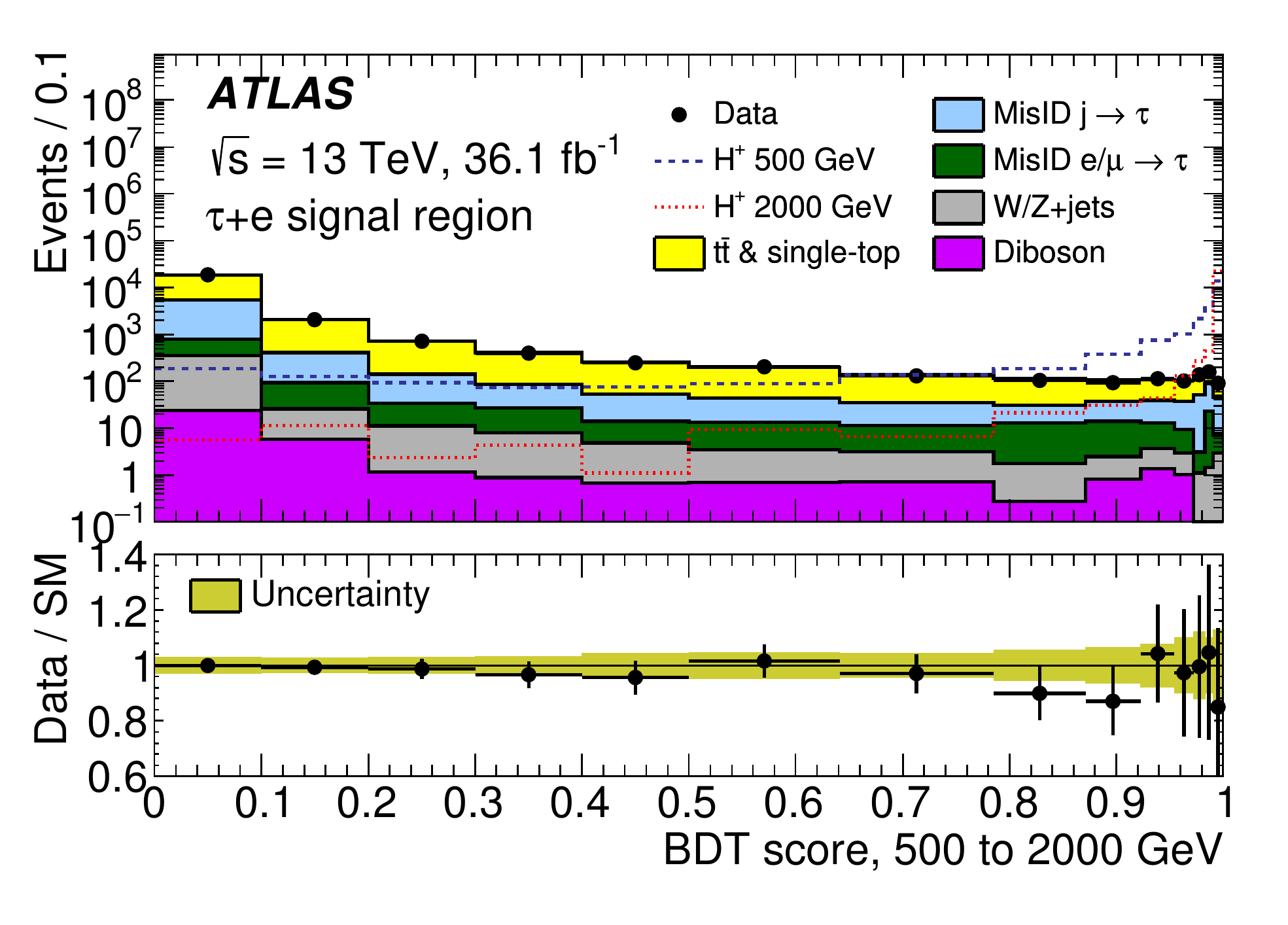}}
\caption{\label{fig:taulepBDTscoreSR1_body} BDT score distributions in the
signal region of the $\tauhadvis$+electron sub-channel, in the five mass ranges used for the
BDT trainings, after a fit to the data with the background-only hypothesis. 
The lower panel of each plot shows the ratio of data to the SM background prediction.  The uncertainty bands include all statistical and systematic uncertainties. 
The normalisation of the signal (shown for illustration) corresponds to the integral of the background.}
\end{figure}

\begin{figure}
\centering
\subfloat[\label{fig:taulepBDTscoreSR2_body_a}]{\includegraphics[width=0.45\linewidth]{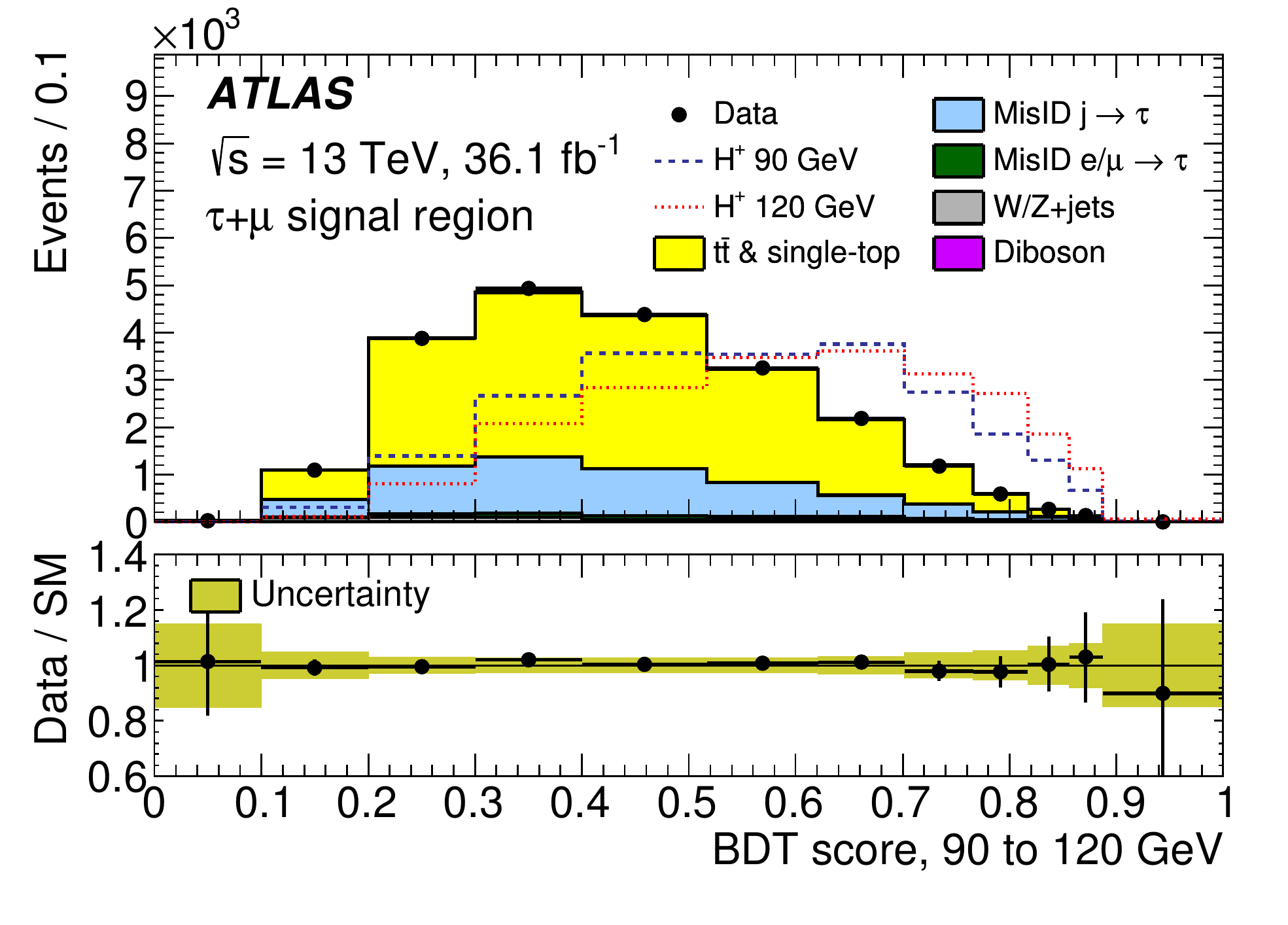}}
\subfloat[\label{fig:taulepBDTscoreSR2_body_b}]{\includegraphics[width=0.45\linewidth]{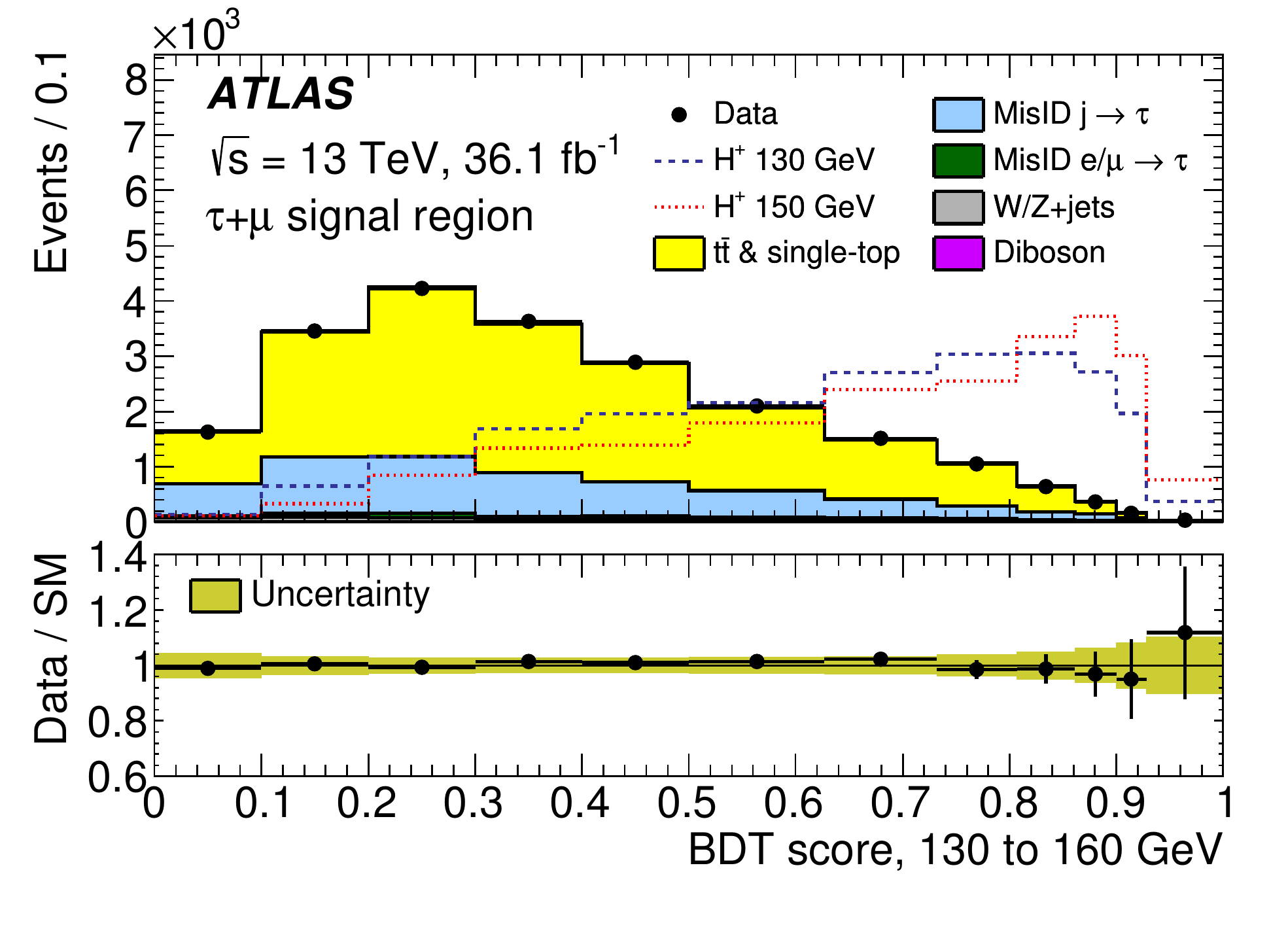}} \\
\subfloat[\label{fig:taulepBDTscoreSR2_body_c}]{\includegraphics[width=0.45\linewidth]{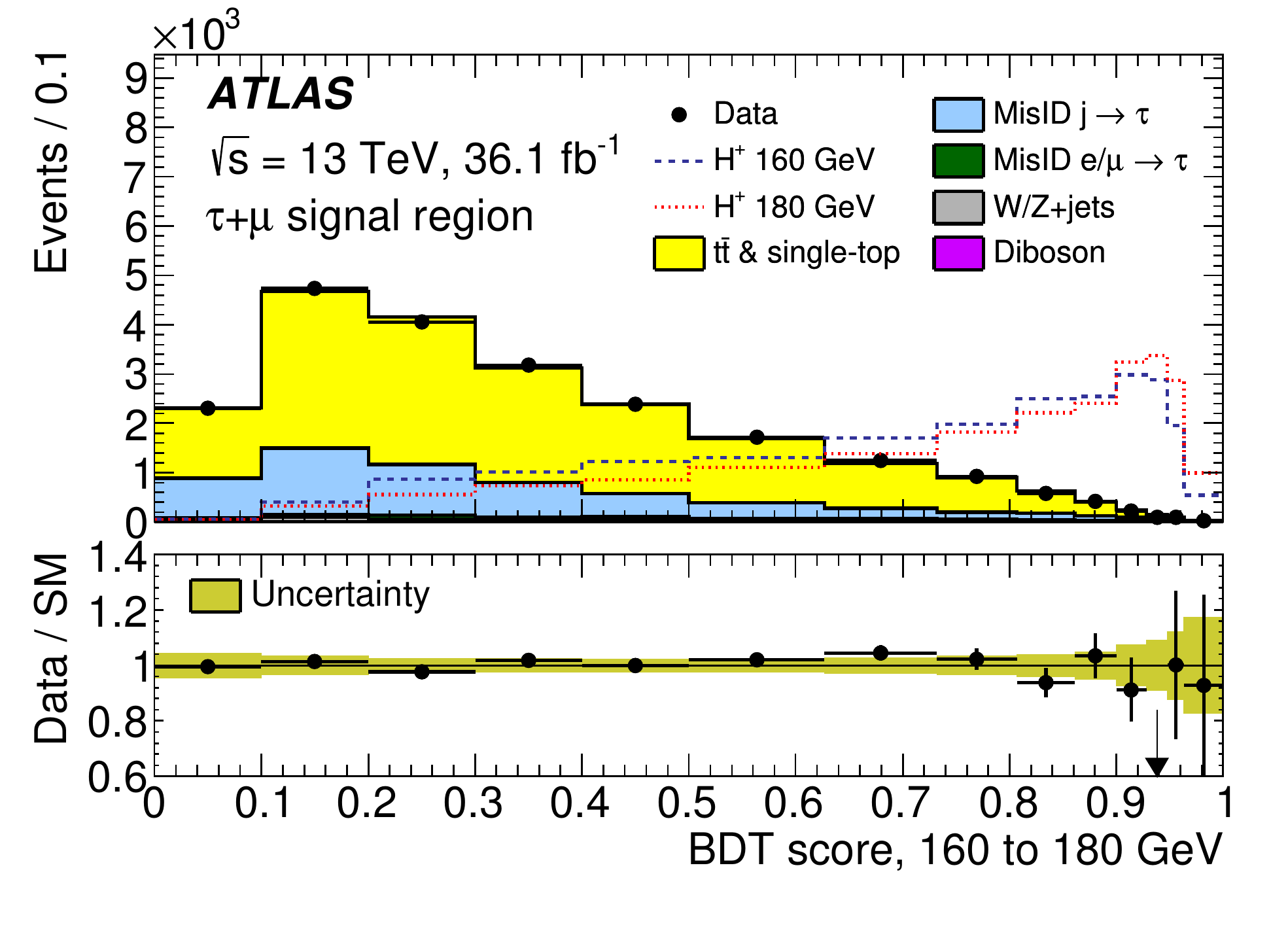}}
\subfloat[\label{fig:taulepBDTscoreSR2_body_d}]{\includegraphics[width=0.45\linewidth]{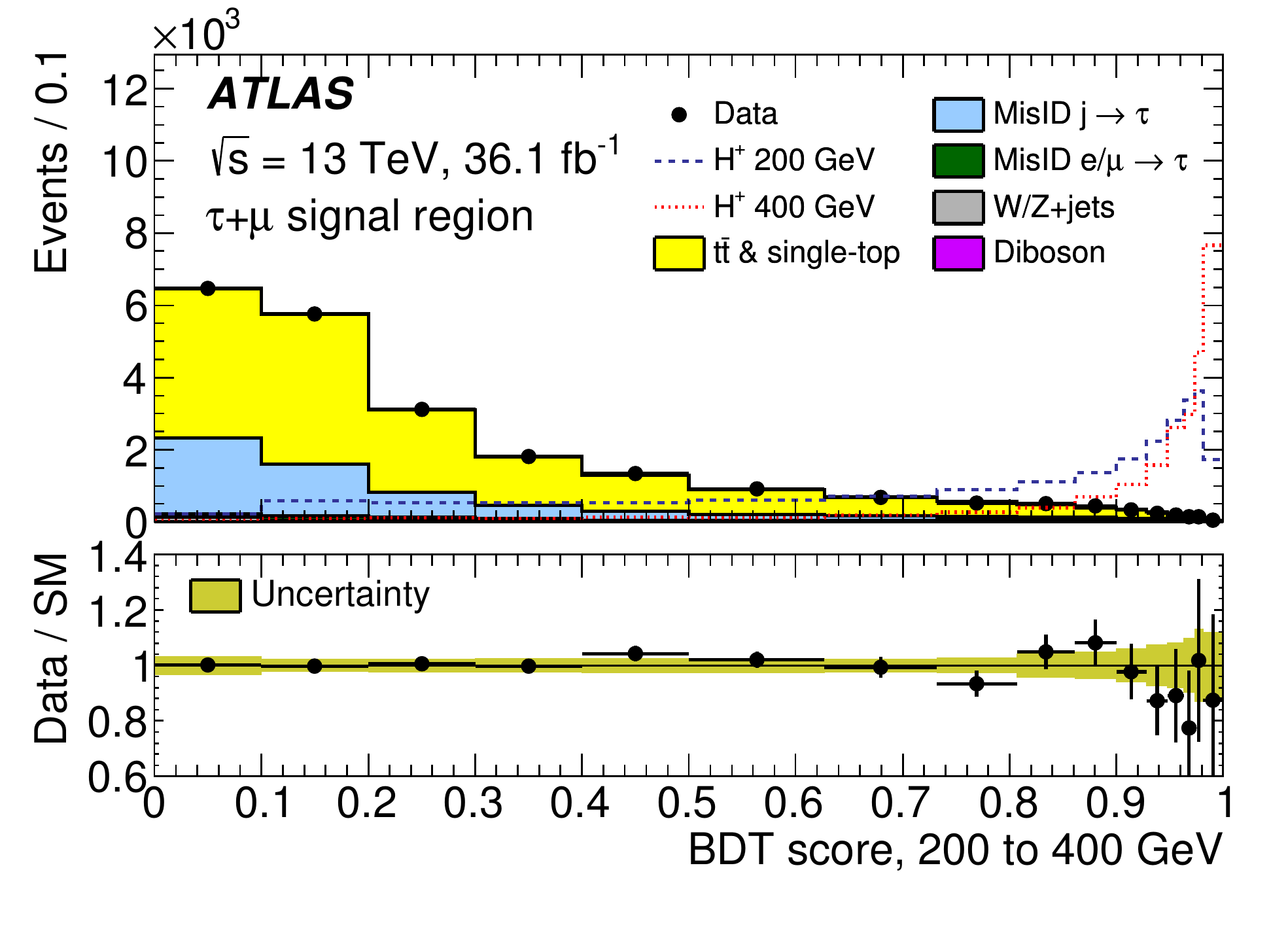}} \\
\subfloat[\label{fig:taulepBDTscoreSR2_body_e}]{\includegraphics[width=0.45\linewidth]{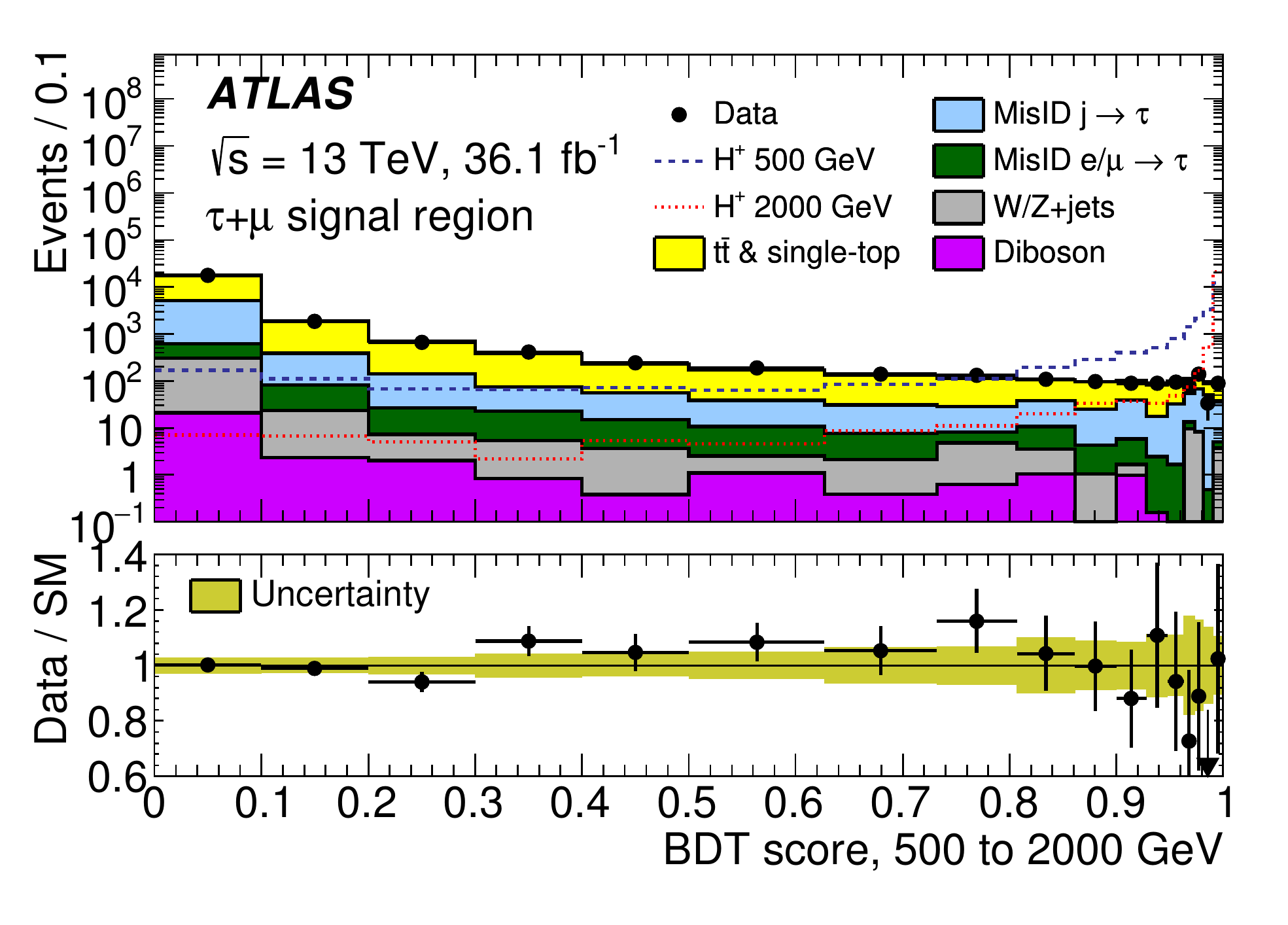}}
\caption{\label{fig:taulepBDTscoreSR2_body} BDT score distributions in the
signal region of the $\tauhadvis$+muon sub-channel, in the five mass ranges used for the
BDT trainings, after a fit to the data with the background-only hypothesis. 
The lower panel of each plot shows the ratio of data to the SM background prediction.  The uncertainty bands include all statistical and systematic uncertainties. 
The normalisation of the signal (shown for illustration) corresponds to the integral of the background.}
\end{figure}

The BDT score distributions in the five charged Higgs boson mass ranges considered in the 
analysis are shown in Figure~\ref{fig:taujets_SR_BDTscores_body} for the signal region of 
the $\tauhadvis$+jets channel, as well as in Figures~\ref{fig:taulepBDTscoreSR1_body} 
and~\ref{fig:taulepBDTscoreSR2_body} for the $\tauhadvis$+electron and $\tauhadvis$+muon 
sub-channels, respectively. All plots are obtained after the statistical fitting procedure 
with the background-only hypothesis. The binning shown in the figures is also used 
in the statistical analysis.

\begin{figure}
\centering
\subfloat[\label{fig:main:limits_a}]{\includegraphics[width=0.5\linewidth]{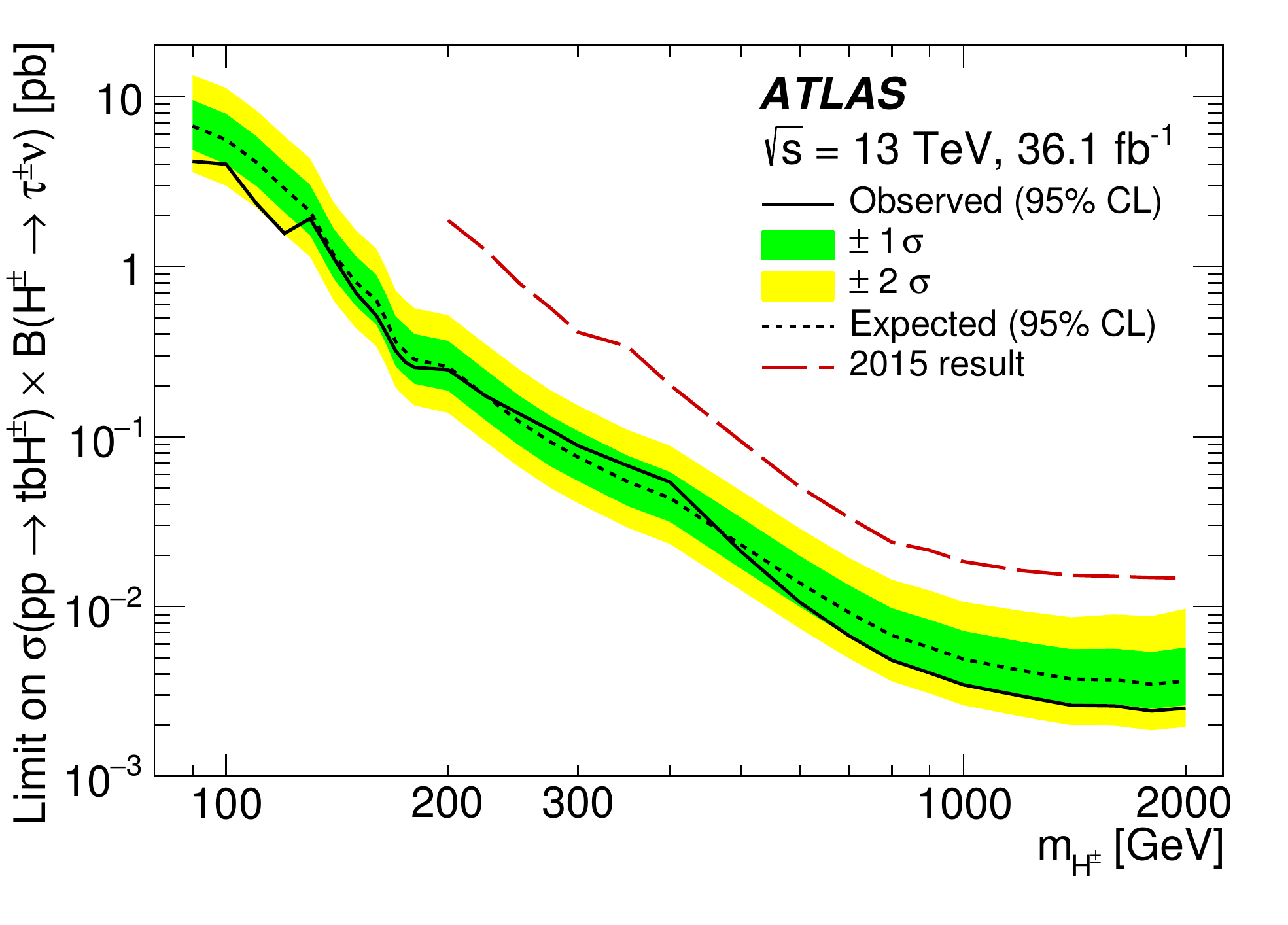}}
\subfloat[\label{fig:main:limits_b}]{\includegraphics[width=0.5\linewidth]{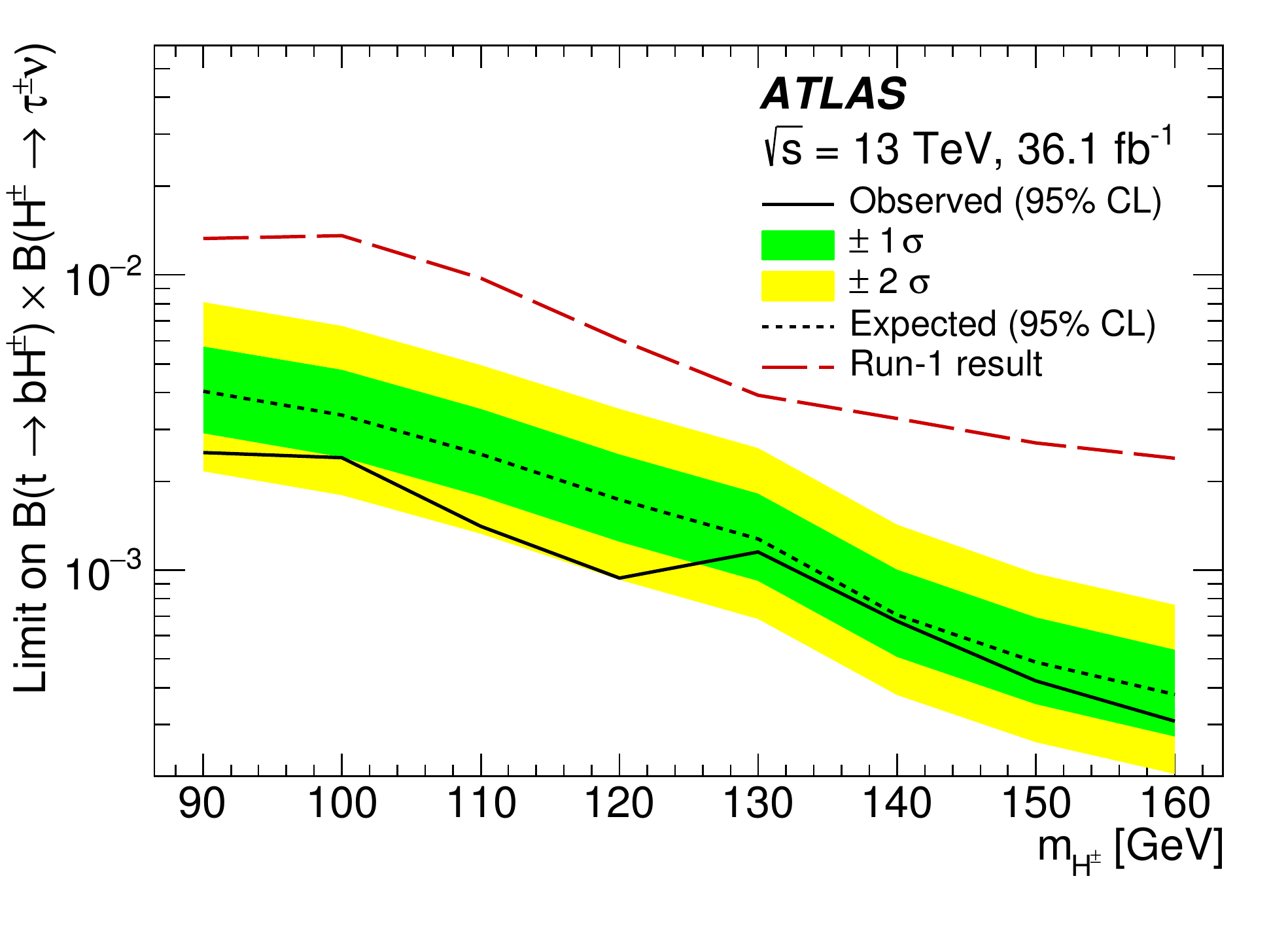}}
\caption{
Observed and expected 95\% CL exclusion limits on (a) $\sigma(pp\to tbH^+)\times \mathrm{\cal{B}}(H^+ \to \tau \nu)$ 
and (b) $\mathrm{\cal{B}}(t\to bH^+)\times \mathrm{\cal{B}}(H^+ \to \tau \nu)$ as a function of the charged Higgs 
boson mass in 36.1~\invfb\ of $pp$ collision data at $\sqrt{s} = 13$~\TeV, after combination of the \tauhadvis+jets 
and \tauhadvis+lepton channels. In the case of the expected limits, one- and two-standard-deviation 
uncertainty bands are also shown. As a comparison, the observed exclusion limits on 
$\mathrm{\cal{B}}(t\to bH^+)\times \mathrm{\cal{B}}(H^+ \to \tau \nu)$
obtained with the Run-1 data at $\rts = \unit[8]{\TeV}$~\cite{Aad:2014kga} and on 
$\sigma(pp\to tbH^+)\times \mathrm{\cal{B}}(H^+ \to \tau \nu)$
obtained with the dataset collected in 2015 at $\rts = \unit[13]{\TeV}$~\cite{Aaboud:2016dig} are also shown.
}
\label{fig:main:limits}
\end{figure}

The data are found to be consistent with the background-only hypothesis (the smallest $p_0$-value is
0.3 around 350~\GeV). Exclusion limits are set at the 95\% confidence level (CL), by using the 
CL$_\text{s}$ procedure~\cite{Read:2002hq}, on 
$\sigma(pp \to tbH^+) \times \mathrm{\cal{B}}(H^+ \to \tau\nu)$ for the full mass range investigated, 
as well as on $\mathrm{\cal{B}}(t\to bH^+)\times \mathrm{\cal{B}}(H^+ \to \tau \nu)$ in the low \Hp\ mass range. 
Figure~\ref{fig:main:limits} shows the expected and observed exclusion limits
as a function of the \Hp\ mass hypothesis. The observed limits range from
4.2~pb to 2.5~fb over the mass range considered in this search. The limits are interpolated between the
\Hp\ mass regions which are tested explicitly. The bias in the expected limits from this interpolation is found 
to be smaller than the statistical uncertainty. For the mass range between 90 and 160~\GeV, the limits 
on $\sigma(pp \to tb\Hp) \times \mathrm{\cal{B}}(\Hp \to \tau\nu)$ translate into observed limits between
0.25\% and 0.031\% for the branching fraction $\mathrm{\cal{B}}(t\to bH^+) \times \mathrm{\cal{B}}(H^+ \to \tau\nu)$ 
if one assumes that the production cross-section is equal to that of \ttbar\ pairs.

The impact from the various sources of systematic uncertainty is estimated by comparing the expected 
95\% CL limits on $\sigma(pp \to tbH^+) \times \mathrm{\cal{B}}(H^+ \to \tau\nu)$ when taking only statistical 
uncertainties into account and those obtained when a certain set of systematic uncertainties is added in 
the limit-setting procedure, as summarised in Table~\ref{tab:impact_sys} for \Hp\ masses of 170~\GeV\ and 1000~\GeV.

\begin{figure}[htp]
  \centering
  \includegraphics[width=0.5\textwidth]{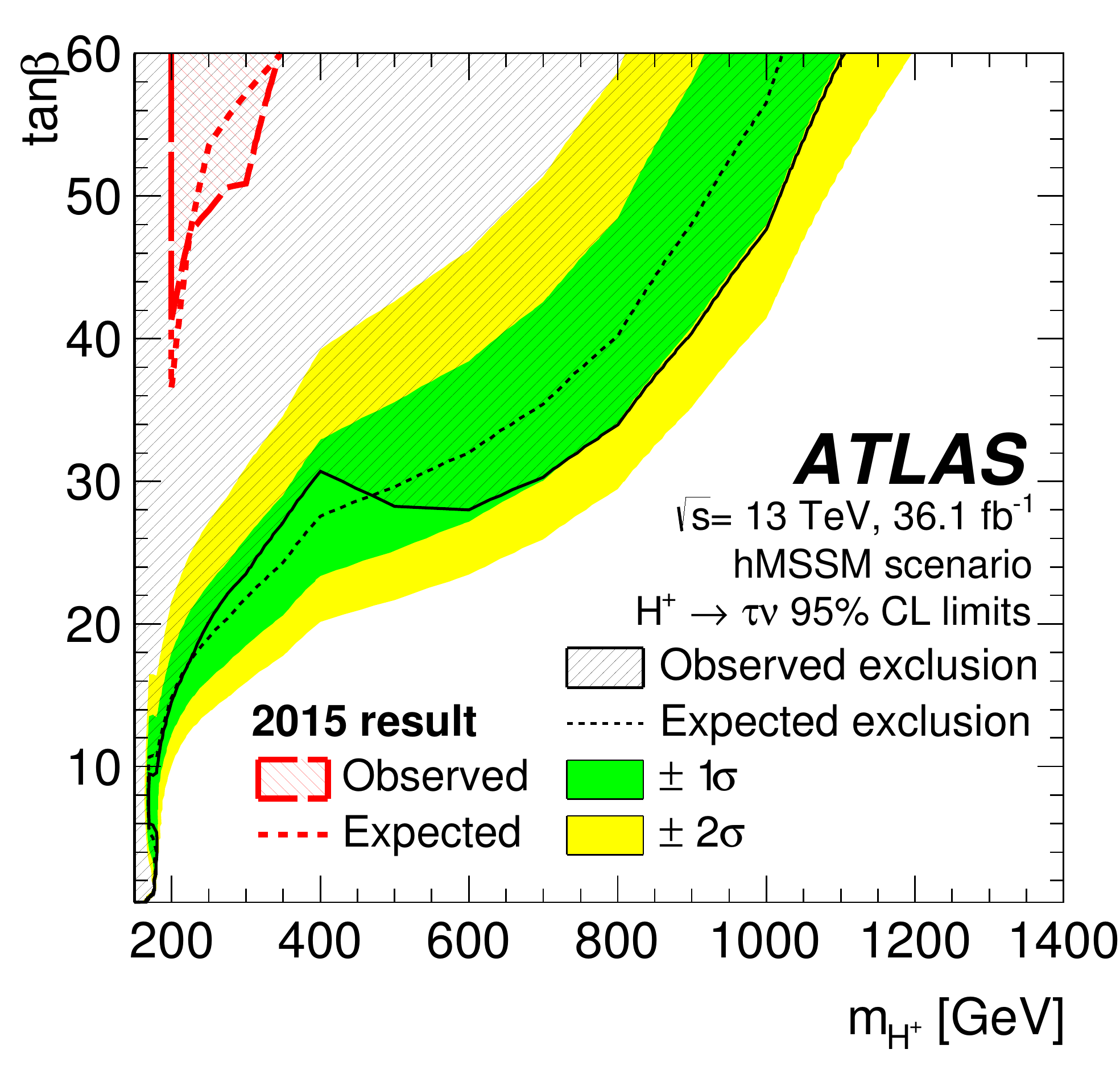}
  \caption{95\% CL exclusion limits on $\tan\beta$ as a function of $m_{\Hp}$,
    shown in the context of the hMSSM, for the regions in which theoretical
    predictions are available $(0.5 \leq \tan\beta \leq 60)$. In the case of the 
    expected limits, one- and two-standard-deviation uncertainty bands are also shown.  
    As a comparison, the red curves near the upper-left corner show the observed 
    and expected exclusion limits based on the dataset of 3.2~\invfb\ collected 
    in 2015 at $\rts = \unit[13]{\TeV}$~\cite{Aaboud:2016dig}.
}
  \label{fig:limits-hmssm}
\end{figure}

Figure~\ref{fig:limits-hmssm}
shows the 95\% CL exclusion limits on $\tan\beta$ as a function of the charged
Higgs boson mass in the context of the hMSSM scenario. All $\tan\beta$ values
are excluded for $m_{H^+} \lesssim 160~\GeV$. At $\tan\beta = 60$,
above which no reliable theoretical calculations exist, the charged Higgs boson
mass range up to 1100~\GeV\ is excluded, hence significantly improving on the limits 
based on the dataset collected in 2015, which corresponds to an
integrated luminosity of 3.2~\invfb.

\begin{table}
\begin{center}
\caption{\label{tab:event_yields_SR_taujets} Expected event yields for the backgrounds and a hypothetical
$\Hp$ signal after applying all $\tauhadvis$+jets selection criteria, and comparison with \LUMI\ of data. All yields
are evaluated prior to using the multivariate discriminant and applying the statistical fitting procedure.
The values shown for the signal assume a charged Higgs boson mass of 170~\GeV\ and 1000~\GeV, 
with a cross-section times branching fraction $\sigma(pp \to tb\Hp) \times \mathrm{\cal{B}}(\Hp \to \tau\nu)$ 
corresponding to $\tan\beta=40$ in the hMSSM benchmark scenario. 
Statistical and systematic uncertainties are quoted, respectively.\\}
\begin{tabular}{l|l}
Sample & Event yields $\tauhadvis$+jets          \\

\hline
True $\tau_{\text{had}}$  & \\
 \hspace*{5mm}$t\bar{t}$ & $\phantom{0}6900\phantom{.00}\pm\phantom{0}60\phantom{.00}\pm1800\phantom{.00}$    \\
 \hspace*{5mm}Single-top-quark  & $\phantom{00}750\phantom{.00}\pm\phantom{0}20\phantom{.00}\pm\phantom{0}100\phantom{.00}$ \\
 \hspace*{5mm}$W \to \tau\nu$  & $\phantom{0}1050\phantom{.00}\pm\phantom{0}30\phantom{.00}\pm\phantom{0}180\phantom{.00}$  \\
 \hspace*{5mm}$Z \to \tau\tau$  & $\phantom{000}84\phantom{.00}\pm\phantom{0}42\phantom{.00}\pm\phantom{00}28\phantom{.00}$   \\
 \hspace*{5mm}Diboson ($WW, WZ, ZZ$) & $\phantom{000}63.2\phantom{0}\pm\phantom{00}4.6\phantom{0}\pm\phantom{000}7.2\phantom{0}$  \\
Misidentified $e,\,\mu \to \tauhadvis$   & $\phantom{00}265\phantom{.00}\pm\phantom{0}12\phantom{.00}\pm\phantom{00}35\phantom{.00}$  \\
Misidentified $\mbox{jet} \to \tauhadvis$ & $\phantom{0}2370\phantom{.00}\pm\phantom{0}20\phantom{.00}\pm \phantom{0}260\phantom{.00}$ \\
\hline
All backgrounds   & $11500\phantom{.00}\pm\phantom{0}80\phantom{.00}\pm1800\phantom{.00}$  \\
\hline
$H^+$ $\phantom{0}$(170~\GeV), hMSSM $\tan\beta=40$ & $\phantom{0}1400\phantom{.00}\pm\phantom{0}10\phantom{.00}\pm\phantom{0}170\phantom{.00}$  \\
$H^+$ (1000~\GeV), hMSSM $\tan\beta=40$ & $\phantom{000}10.33\pm\phantom{00}0.06\pm\phantom{000}0.78$  \\
\hline
Data         &  11021   \\
\hline
\end{tabular}
\end{center}
\end{table}

\begin{table}
\begin{center}
\caption{\label{tab:event_yields_SR_taulep} Expected event yields for the backgrounds and a hypothetical
$\Hp$ signal after applying all $\tauhadvis$+lepton selection criteria, and comparison with \LUMI\ of data. All yields
are evaluated prior to using the multivariate discriminant and applying the statistical fitting procedure. The
values shown for the signal assume a charged Higgs boson mass of 170~\GeV\ and 1000~\GeV, 
with a cross-section times branching fraction $\sigma(pp \to tb\Hp) \times \mathrm{\cal{B}}(\Hp \to \tau\nu)$ 
corresponding to $\tan\beta=40$ in the hMSSM benchmark scenario. 
Statistical and systematic uncertainties are quoted, respectively.\\}
\begin{tabular}{l|l|l}
Sample &   Event yields $\tauhadvis$+electron  & Event yields $\tauhadvis$+muon \\
\hline
True $\tau_{\text{had}}$  & \\
 \hspace*{5mm}$t\bar{t}$ & $16000\phantom{.00}\pm\phantom{0}80\phantom{.00}\pm2500$ & $14600\phantom{.00}\pm\phantom{0}80\phantom{.00}\pm2400$  \\
 \hspace*{5mm}Single-top-quark  & $\phantom{0}1260\phantom{.00}\pm\phantom{0}20\phantom{.00}\pm\phantom{0}110$ & $\phantom{0}1260\phantom{.00}\pm\phantom{0}20\phantom{.00}\pm\phantom{0}110$  \\
 \hspace*{5mm}$Z \to \tau\tau$  & $\phantom{00}433\phantom{.00}\pm\phantom{0}27\phantom{.00}\pm\phantom{00}80$  & $\phantom{00}352\phantom{.00}\pm\phantom{0}48\phantom{.00}\pm\phantom{00}43$ \\
 \hspace*{5mm}Diboson ($WW, WZ, ZZ$) &  $\phantom{000}39.3\phantom{0}\pm\phantom{00}2.1\phantom{0}\pm\phantom{000}4.5$ & $\phantom{000}32.3\phantom{0}\pm\phantom{00}1.7\phantom{0}\pm\phantom{000}3.6$ \\
Misidentified $e,\,\mu \to \tauhadvis$   &   $\phantom{00}626\phantom{.00}\pm\phantom{0}27\phantom{.00}\pm\phantom{00}59$ & $\phantom{00}454\phantom{.00}\pm\phantom{0}16\phantom{.00}\pm\phantom{00}27$ \\
Misidentified $\mbox{jet} \to \tauhadvis$ &   $\phantom{0}5640\phantom{.00}\pm\phantom{0}40\phantom{.00}\pm\phantom{0}450$ & $\phantom{0}5460\phantom{.00}\pm\phantom{0}40\phantom{.00}\pm\phantom{0}410$ \\
\hline
All backgrounds   &  $24000\phantom{.00}\pm100\phantom{.00}\pm2600$ & $22200\phantom{.00}\pm100\phantom{.00}\pm2500$ \\
\hline
$H^+$ $\phantom{0}$(170~\GeV), hMSSM $\tan\beta=40$ & $\phantom{00}850\phantom{.00}\pm\phantom{0}12\phantom{.00}\pm\phantom{00}65$ & $\phantom{00}852\phantom{.00}\pm\phantom{0}11\phantom{.00}\pm\phantom{00}66$ \\
$H^+$ (1000~\GeV), hMSSM $\tan\beta=40$ &   $\phantom{0000}0.82\pm\phantom{00}0.02\pm\phantom{000}0.07$ & $\phantom{0000}1.05\pm\phantom{00}0.02\pm\phantom{000}0.09$ \\
\hline
Data         &     22645 &  21419  \\
\hline
\end{tabular}
\end{center}
\end{table}

\begin{table}
\begin{center}
\caption{\label{tab:impact_sys} Impact of systematic uncertainties on the expected 95\% CL
limit on $\sigma(pp \to tbH^+) \times \mathrm{\cal{B}}(H^+ \to \tau\nu)$, for two $H^+$ mass 
hypotheses: 170~\GeV\ and 1000~\GeV. The impact is obtained by comparing the expected limit considering only 
statistical uncertainties (stat.\,only) with the expected limit when a certain set of systematic uncertainties is 
added in the limit-setting procedure. 
In the absence of correlations and assuming Gaussian uncertainties, the row ``All'' would be obtained by 
summing in quadrature (linearly) the individual contributions of the systematic uncertainties if these were 
much larger (smaller) than the statistical uncertainties.\\}
\begin{tabular}{l|cc}
Source of systematic & \multicolumn{2}{c}{Impact on the expected limit (stat.\,only) in \%} \\
uncertainty & $m_{H^{+}} = 170~\GeV$ & $m_{H^{+}} = 1000~\GeV$ \\
\hline
Experimental & & \\
\hspace*{5mm} luminosity  &  $\phantom{<}$$\phantom{0}2.9$   &  $\phantom{<}$$\phantom{0}0.2$ \\
\hspace*{5mm} trigger  &  $\phantom{<}$$\phantom{0}1.3$  &  $\phantom{0}$$<$$0.1$ \\
\hspace*{5mm} \tauhadvis\ & $\phantom{<}$$14.6$ & $\phantom{<}$$\phantom{0}$0.3\\
\hspace*{5mm} jet &  $\phantom{<}$$16.9$   &  $\phantom{<}$$\phantom{0}0.2$ \\
\hspace*{5mm} electron & $\phantom{<}$$10.1$ & $\phantom{<}$$\phantom{0}0.1$\\
\hspace*{5mm} muon & $\phantom{<}$$\phantom{0}1.1$ & $\phantom{0}$$<$$0.1$\\
\hspace*{5mm} $\met$  &  $\phantom{<}$$\phantom{0}9.9$ & $\phantom{0}$$<$$0.1$ \\
\hline
Fake-factor method &  $\phantom{<}$$20.3$   &  $\phantom{<}$$\phantom{0}2.7$ \\
$\Upsilon$ modelling &  $\phantom{<}$$\phantom{0}0.8$ & $\phantom{<}$$\phantom{0}-$\\
\hline
Signal and background models & & \\
\hspace*{5mm} $t\bar{t}$ modelling   &  $\phantom{<}$$\phantom{0}6.3$   &  $\phantom{<}$$\phantom{0}0.1$ \\
\hspace*{5mm} $W/Z$+jets modelling   &  $\phantom{<}$$\phantom{0}1.1$ & $\phantom{0}$$<$$0.1$ \\
\hspace*{5mm} cross-sections $(W/Z/VV/t)$ & $\phantom{<}$$\phantom{0}9.6$ & $\phantom{<}$$\phantom{0}0.4$\\
\hspace*{5mm} $H^{+}$ signal modelling  &  $\phantom{<}$$\phantom{0}2.5$   &  $\phantom{<}$$\phantom{0}6.4$ \\
\hline
All  &  $\phantom{<}52.1$ & $\phantom{<}13.8$ \\
\hline
\end{tabular}
\end{center}
\end{table}

\FloatBarrier

\section{Conclusion}
\label{sec:conclusion}
A search for charged Higgs bosons produced either in top-quark decays or
in association with a top-quark, and subsequently decaying via $\Hp \to \tau \nu$, 
is performed in the $\tau$+jets and $\tau$+lepton channels, according to the 
hadronic or semileptonic decay of the top quark produced together with \Hp.
The dataset contains \LUMI\ of $pp$ collisions at $\sqrt{s} = 13$~\TeV, 
recorded with the ATLAS detector at the LHC. The data are 
found to be in agreement with the background-only hypothesis. Upper limits 
at the 95\% confidence level are set on the \Hp\ production cross-section times 
the branching fraction $\mathrm{\cal{B}}(H^+ \to \tau \nu)$ between 4.2~pb and 2.5~fb 
for a charged Higgs boson mass range of 90--2000~\GeV, corresponding to upper limits between 0.25\% 
and 0.031\% for the branching fraction $\mathrm{\cal{B}}(t\to bH^+) \times \mathrm{\cal{B}}(H^+ \to \tau\nu)$ 
in the mass range 90--160~\GeV. These exclusion limits are about 5--7 times more stringent than 
those obtained by ATLAS with 3.2~\invfb\ of 13~\TeV\ data for \Hp\ masses above 200~\GeV~\cite{Aaboud:2016dig} 
and with Run-1 data in the \Hp\ mass range 90--160~\GeV~\cite{Aad:2014kga}. In the intermediate-mass 
region where $m_{H^+} \simeq m_{\text{top}}$, accurate theoretical predictions recently became available, 
allowing a dedicated comparison of the \Hp\ models with data near the top-quark mass. In the context 
of the hMSSM scenario, all $\tan\beta$ values are excluded for $m_{\Hp} \lesssim 160~\GeV$. The $\Hp$ 
mass range up to 1100~\GeV\ is excluded at $\tan\beta=60$.

\section*{Acknowledgements}

We thank CERN for the very successful operation of the LHC, as well as the
support staff from our institutions without whom ATLAS could not be
operated efficiently.

We acknowledge the support of ANPCyT, Argentina; YerPhI, Armenia; ARC, Australia; BMWFW and FWF, Austria; ANAS, Azerbaijan; SSTC, Belarus; CNPq and FAPESP, Brazil; NSERC, NRC and CFI, Canada; CERN; CONICYT, Chile; CAS, MOST and NSFC, China; COLCIENCIAS, Colombia; MSMT CR, MPO CR and VSC CR, Czech Republic; DNRF and DNSRC, Denmark; IN2P3-CNRS, CEA-DRF/IRFU, France; SRNSFG, Georgia; BMBF, HGF, and MPG, Germany; GSRT, Greece; RGC, Hong Kong SAR, China; ISF, I-CORE and Benoziyo Center, Israel; INFN, Italy; MEXT and JSPS, Japan; CNRST, Morocco; NWO, Netherlands; RCN, Norway; MNiSW and NCN, Poland; FCT, Portugal; MNE/IFA, Romania; MES of Russia and NRC KI, Russian Federation; JINR; MESTD, Serbia; MSSR, Slovakia; ARRS and MIZ\v{S}, Slovenia; DST/NRF, South Africa; MINECO, Spain; SRC and Wallenberg Foundation, Sweden; SERI, SNSF and Cantons of Bern and Geneva, Switzerland; MOST, Taiwan; TAEK, Turkey; STFC, United Kingdom; DOE and NSF, United States of America. In addition, individual groups and members have received support from BCKDF, the Canada Council, CANARIE, CRC, Compute Canada, FQRNT, and the Ontario Innovation Trust, Canada; EPLANET, ERC, ERDF, FP7, Horizon 2020 and Marie Sk{\l}odowska-Curie Actions, European Union; Investissements d'Avenir Labex and Idex, ANR, R{\'e}gion Auvergne and Fondation Partager le Savoir, France; DFG and AvH Foundation, Germany; Herakleitos, Thales and Aristeia programmes co-financed by EU-ESF and the Greek NSRF; BSF, GIF and Minerva, Israel; BRF, Norway; CERCA Programme Generalitat de Catalunya, Generalitat Valenciana, Spain; the Royal Society and Leverhulme Trust, United Kingdom.

The crucial computing support from all WLCG partners is acknowledged gratefully, in particular from CERN, the ATLAS Tier-1 facilities at TRIUMF (Canada), NDGF (Denmark, Norway, Sweden), CC-IN2P3 (France), KIT/GridKA (Germany), INFN-CNAF (Italy), NL-T1 (Netherlands), PIC (Spain), ASGC (Taiwan), RAL (UK) and BNL (USA), the Tier-2 facilities worldwide and large non-WLCG resource providers. Major contributors of computing resources are listed in Ref.~\cite{ATL-GEN-PUB-2016-002}.

\clearpage

\printbibliography
\clearpage

 
\begin{flushleft}
{\Large The ATLAS Collaboration}

\bigskip

M.~Aaboud$^\textrm{\scriptsize 34d}$,    
G.~Aad$^\textrm{\scriptsize 99}$,    
B.~Abbott$^\textrm{\scriptsize 124}$,    
O.~Abdinov$^\textrm{\scriptsize 13,*}$,    
B.~Abeloos$^\textrm{\scriptsize 128}$,    
D.K.~Abhayasinghe$^\textrm{\scriptsize 91}$,    
S.H.~Abidi$^\textrm{\scriptsize 164}$,    
O.S.~AbouZeid$^\textrm{\scriptsize 39}$,    
N.L.~Abraham$^\textrm{\scriptsize 153}$,    
H.~Abramowicz$^\textrm{\scriptsize 158}$
% ...
\bigskip
\\

% ...
$^{13}$Institute of Physics, Azerbaijan Academy of Sciences, Baku; Azerbaijan.\\
% ...
$^{34}$$^{(a)}$Facult\'e des Sciences Ain Chock, R\'eseau Universitaire de Physique des Hautes Energies - Universit\'e Hassan II, Casablanca;$^{(b)}$Centre National de l'Energie des Sciences Techniques Nucleaires (CNESTEN), Rabat;$^{(c)}$Facult\'e des Sciences Semlalia, Universit\'e Cadi Ayyad, LPHEA-Marrakech;$^{(d)}$Facult\'e des Sciences, Universit\'e Mohamed Premier and LPTPM, Oujda;$^{(e)}$Facult\'e des sciences, Universit\'e Mohammed V, Rabat; Morocco.\\
% ...
$^{39}$Niels Bohr Institute, University of Copenhagen, Copenhagen; Denmark.\\
% ...
$^{91}$Department of Physics, Royal Holloway University of London, Egham; United Kingdom.\\
% ...
$^{99}$CPPM, Aix-Marseille Universit\'e, CNRS/IN2P3, Marseille; France.\\
% ...
$^{124}$Homer L. Dodge Department of Physics and Astronomy, University of Oklahoma, Norman OK; United States of America.\\
% ...
$^{128}$LAL, Universit\'e Paris-Sud, CNRS/IN2P3, Universit\'e Paris-Saclay, Orsay; France.\\
% ...
$^{153}$Department of Physics and Astronomy, University of Sussex, Brighton; United Kingdom.\\
% ...
$^{158}$Raymond and Beverly Sackler School of Physics and Astronomy, Tel Aviv University, Tel Aviv; Israel.\\
% ...
$^{164}$Department of Physics, University of Toronto, Toronto ON; Canada.\\
% ...
$^{a}$ Also at Borough of Manhattan Community College, City University of New York, NY; United States of America.\\
$^{b}$ Also at Centre for High Performance Computing, CSIR Campus, Rosebank, Cape Town; South Africa.\\
$^{c}$ Also at CERN, Geneva; Switzerland.\\
$^{d}$ Also at CPPM, Aix-Marseille Universit\'e, CNRS/IN2P3, Marseille; France.\\
$^{e}$ Also at D\'epartement de Physique Nucl\'eaire et Corpusculaire, Universit\'e de Gen\`eve, Gen\`eve; Switzerland.\\
% ...
$^{*}$ Deceased

\end{flushleft}


\end{document}